\RequirePackage{fix-cm}
\RequirePackage{amsmath}
\documentclass[a4paper]{article}
\usepackage{marginnote} 
\usepackage{wallpaper} 
\usepackage{amsmath} 
\usepackage{amssymb} 
\usepackage{xcolor} 
\usepackage{verbatim}	
\usepackage{pdfpages}	
\usepackage{graphicx}
\usepackage{tabularx}
\usepackage{multirow}
\usepackage{enumitem}
\usepackage{subfiles}
\usepackage[T1]{fontenc}
\usepackage[labelfont=bf]{caption}
\usepackage[utf8x]{inputenc}
\usepackage{booktabs}
\usepackage[title,titletoc,toc]{appendix}
\usepackage{lscape}
\usepackage{eurosym}

\usepackage{lmodern}
\usepackage[T1]{fontenc}
\usepackage{blindtext}
\usepackage{hyperref}

\newcommand{\apjl}{The Astrophysical Journal Letters}
\newcommand{\apjs}{The Astrophysical Journal Supplement}

\newcommand{\aap}{A\&A}  

\usepackage{authblk}

\begin{document}

\title{Exoplanet Spectroscopy and Photometry with the Twinkle Space Telescope}

\author[1]{Billy Edwards}
\author[2]{Malena Rice}
\author[1,3]{Tiziano Zingales}
\author[4,1]{Marcell Tessenyi}
\author[1]{Ingo Waldmann}
\author[1,4]{Giovanna Tinetti}
\author[5,6]{Enzo Pascale}
\author[6]{Subhajit Sarkar}

\affil[1]{Department of Physics and Astronomy, University College London, Gower Street, London WC1E 6BT, UK. E-mail: billy.edwards.16@ucl.ac.uk}
\affil[2]{Yale University, Department of Astronomy, Steinbach Hall, New Haven, CT 06511, USA}
\affil[3]{INAF, Osservatorio Astronomico di Palermo, 90134 Palermo, Italy}
\affil[4]{Blue Skies Space Ltd., 69 Wilson Street, London, UK}
\affil[5]{Dipartimento di Fisica, La Sapienza Universita di Roma, Piazzale Aldo Moro 2, 00185 Roma, Italy}
\affil[6]{Cardiff University, School of Physics and Astronomy, Queens Buildings, The Parade, Cardiff, CF24 3AA, UK}

\date{}

\maketitle

\begin{abstract}

The Twinkle space telescope has been designed for the characterisation of exoplanets and Solar System objects. Operating in a low Earth, Sun-synchronous orbit, Twinkle is equipped with a 45 cm telescope and visible (0.4 - 1$\mu$m) and infrared (1.3 - 4.5$\mu$m) spectrometers which can be operated simultaneously. Twinkle is a general observatory which will provide on-demand observations of a wide variety of targets within wavelength ranges that are currently not accessible using other space telescopes or accessible only to oversubscribed observatories in the short-term future.

Here we explore the ability of Twinkle’s spectrometers to characterise the currently-known exoplanets. We study the spectral resolution achievable by combining multiple observations for various planetary and stellar types. We also simulate spectral retrievals for some well-known planets (HD 209458 b, GJ 3470 b and 55 Cnc e).

From the exoplanets known today, we find that with a single transit or eclipse, Twinkle could probe 89 planets at low spectral resolution (R \textless 20) as  well as 12 planets at higher resolution (R \textgreater 20) in channel 1 (1.3 - 4.5$\mu$m). With 10 observations, the atmospheres of 144 planets could be characterised with R \textless 20 and 81 at higher resolutions.

Upcoming surveys will reveal thousands of new exoplanets, many of which will be located within Twinkle’s field of regard. TESS in particular is predicted to discover many targets around bright stars which will be suitable for follow-up observations. We include these anticipated planets and find that the number of planets Twinkle could observe in the near infrared in a single transit or eclipse increases to 558 for R \textgreater 20 and 41 at lower resolutions. By stacking 10 transits or eclipses, there are 1185 potential targets for study at R \textless 20 as well as 388 planets at higher resolutions.

The majority of targets are found to be large gaseous planets although by stacking multiple observations smaller planets around bright stars (e.g. 55 Cnc e) could be observed with Twinkle. Photometry and low resolution spectroscopy with Twinkle will be useful to refine planetary, stellar and orbital parameters, monitor stellar activity through time and search for transit time and duration variations (TTVs and TDVs). Refinement of these parameters could be used to in the planning of observations with larger space-based observatories such as JWST and ARIEL. For planets orbiting very bright stars, Twinkle observations at higher spectral resolution will enable us to probe the chemical and thermal properties of an atmosphere. Simultaneous coverage across a wide wavelength range will reduce the degeneracies seen with Hubble and provide access to detections of a wide range molecules. There is the potential to revisit them many times over the mission lifetime to detect variations in cloud cover. 

\end{abstract}

\vspace{5mm}
\textbf{Acknowledgements}

This research has made use of the NASA Exoplanet Archive, which is operated by the California Institute of Technology, under contract with the National Aeronautics and Space Administration under the Exoplanet Exploration Program. Additionally, the Open Exoplanet Catalogue, TEPCat and exoplanet.eu have been utilised as supplementary data sources.

This work has been funded through the ERC Consolidator grant ExoLights (GA 617119) and the STFC grant ST/S002634/1.

\section{Introduction}
As of July 2018, over 3700 exoplanets have been discovered (nearly 3000 of which transit their stars) as well as 2700 Kepler candidates yet to be confirmed as planets. On top of this, in the next few years Gaia is anticipated to discover up to ten thousand Jupiter-sized planets \cite{sozzetti,perryman} whilst TESS is expected to detect thousands of transiting planets of Earth size or larger \cite{ricker,sullivan,barclay}. Additionally, space surveys CHEOPS \cite{broeg} and K2 \cite{howell}, along with ground-based surveys like NGTS \cite{wheatley}, ESPRESSO \cite{pepe}, WASP \cite{pollacco}, HATNet \cite{bakos}, HATSouth \cite{bakos_south}, MEarth \cite{nutzman}, TRAPPIST \cite{jehin}, CARMENES \cite{quirrenbach}, SPIROU \cite{artigau} and HARPS \cite{pepe_harps}, will lead to many more transiting exoplanet detections as well as further characterisation of planetary parameters.

\subsection{Characterisation of Exoplanet Atmospheres from Space}

Whilst many planets have been detected and it is thought that planets are common in our galaxy (e.g. \cite{howard,batalha,cassan,dressing,wright_jup}), our current knowledge of their atmospheric, thermal and compositional characteristics is still very limited. Space telescopes such as Hubble and Spitzer, as well as some ground-based observatories, have provided constraints on these properties for a limited number of targets and, in some cases, have identified the key molecules present in their atmospheres whilst also detecting the presence of clouds and probing the thermal structure (e.g. \cite{brogi,majeau,stevenson,sing,fu,tsiaras_30planets,zhang}). However, the breadth and quality of currently available data is limited by the absence of a dedicated space-based exoplanet spectroscopy mission and therefore progress in this area has been slower than desired.

Spitzer is, along with Hubble, part of NASA’s Great Observatories Program. Launched in 2003, Spitzer carried an Infrared Array Camera (IRAC), an Infrared Spectrograph (IRS) and a Multiband Imaging Photometer (MIPS). The IRS was split over four sub-modules with operational wavelengths of 5.3 - 40$\mu$m \cite{houck} and has not been operational since Spitzer’s helium coolant was depleted in 2009. Since the cool phase of Spitzer’s mission ended, only the IRAC has remained operational, though with reduced capabilities. The Hubble WFC3 currently delivers spectroscopic data at wavelengths shorter than 1.7$\mu$m. Thus, at the time of writing no space telescope capable of infrared spectroscopy beyond 1.7$\mu$m is operational.

The James Webb Space Telescope (JWST) is expected to be launched in March 2021. A Near-Infrared Spectrometer (NIRSpec) and camera (NIRCam) are included within the instrument suite \cite{davila} and thus will provide the infrared capability that is currently missing (0.6 - 5.3$\mu$m and 0.6 - 5.0$\mu$m respectively). Additionally, the Mid-Infrared Instrument (MIRI) covers the wavelength range 5 - 28$\mu$m and is capable of medium resolution spectroscopy whilst the Near-Infrared Imager and Slitless Spectrograph (NIRISS) will cover visible and near infrared wavelengths from 0.6 - 5.0$\mu$m \cite{wright_jwst}. JWST has the capability to view a wide range of exoplanet targets \cite{beichman,greene,louie}. Although a good fraction of JWST observation time is expected to be allocated for exoplanet science (e.g. \cite{cowan,bean}), for a space observatory of this scale, over-subscription is likely to be an issue and not all interesting science cases will necessarily require the sensitivity and accuracy of JWST. 

Further into the future, the ESA ARIEL space mission will conduct a population survey of the atmospheres of $\sim$1000 transiting exoplanets over a wide wavelength range (0.5 - 7.8$\mu$m) followed by a detailed study of a few hundred selected planets. ARIEL aims to deliver a comprehensive catalogue of planetary spectra, which will yield molecular abundances, chemical gradients and atmospheric structures \cite{tinetti_ariel,puig_2018}. ARIEL is expected to be launched in 2028.

\subsection{Twinkle}

The Twinkle Space Mission is a new, fast-track satellite designed for launch in early 2022. It has been conceived for providing faster access to spectroscopic data from exoplanet atmospheres and Solar System bodies, but it is also capable of providing spectra of brown dwarfs, bright stars, nebulae and galaxies. Twinkle is equipped with a visible (0.4 - 1$\mu$m) and infrared (1.3 - 4.5{$\mu$}m) spectrometer (split into two channels at 2.42$\mu$m). Twinkle has been designed with a telescope aperture of 45cm and will operate in a low Earth, Sun-synchronous orbit \cite{savini,jason}.

Twinkle is a general observatory which is being managed by Blue Skies Space Ltd (BSSL). Scientists will be able to purchase telescope time and Twinkle will provide on-demand observations of a wide variety of targets within wavelength ranges that are currently not accessible using other space telescopes or accessible only to oversubscribed observatories in the short-term future. Whilst it is thought that Twinkle will have considerable capabilities for observing Solar System Objects \cite{edwards}, the focus of this paper is to explore its spectrophotometry capabilities to observe a variety of exoplanets in the optical and infrared wavelength regimes.

The Exoplanet Light Visible Spectrometer (ELVIS) incorporated into the Twinkle instrumentation is an adapted version of the Ultraviolet and Visible Spectrometer (UVIS) flown on the ExoMars Trace Gas Orbiter. By modifying the grating and its coatings, the optimum spectral range of the instrument will be changed from 0.2 - 0.6$\mu$m to 0.4 - 1$\mu$m \cite{savini} with a resolving power R$\sim$250. Twinkle’s infrared spectrometers will provide R$\sim$250 for $\lambda$ \textless 2.42$\mu$m and R$\sim$60 for $\lambda$ \textless 2.42$\mu$m \cite{wells}.

\begin{figure}[h!]
  \centering
  \includegraphics[width=1\textwidth]{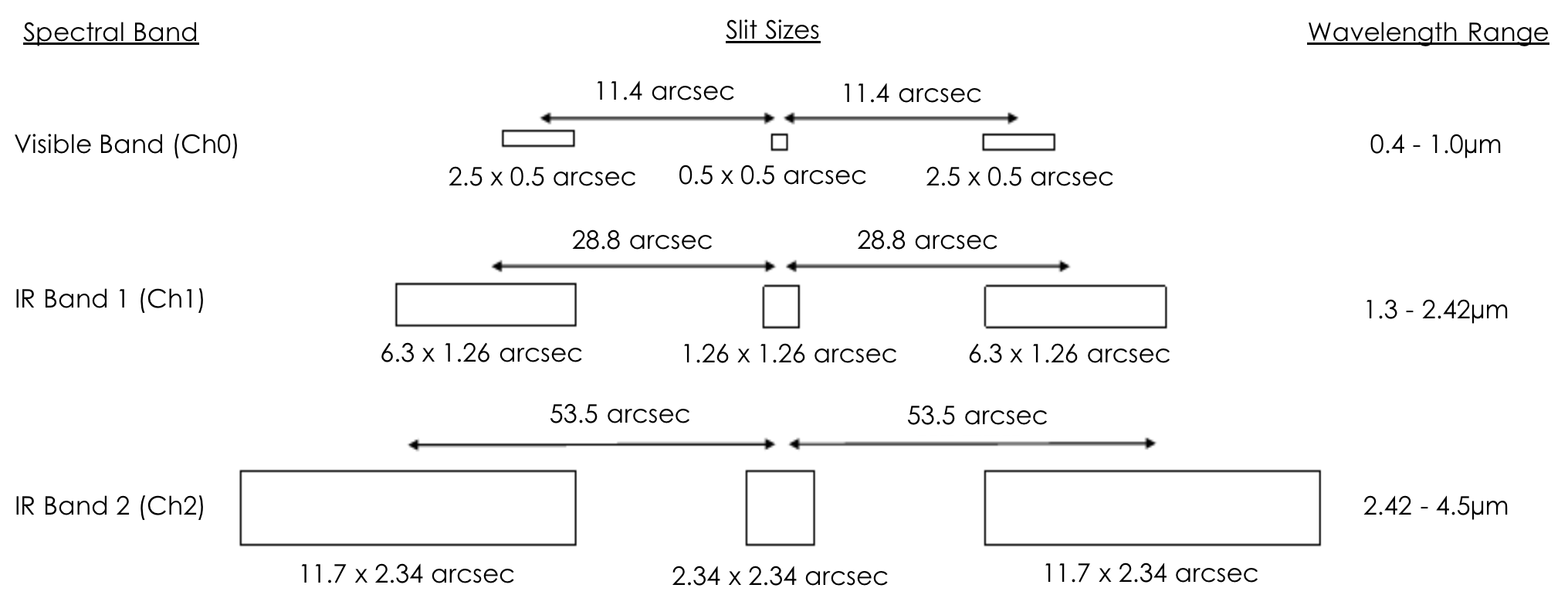}
  \caption{Angular sizes of Twinkle’s star and background slits. Whilst slit sizes are each to scale, the separation between them is not} \label{Slit Sizes}
\end{figure}

The light entering Twinkle’s telescope will be directed through three entrance slits as it is focused onto the instrumentation. Figure \ref{Slit Sizes} shows the expected angular sizes of Twinkle’s slits for each spectral band. Lenses produce several spectra on the detector, with the spectrum from the star slit in the centre and with three spectra from the background slits on either side \cite{wells}. The background slits are utilised to provide spectra of the sky close to the star, permitting subtraction of background light from the planet/star spectra.

For exoplanet observations the cadence will vary depending upon the brightness of the host star. For bright targets, such as 55 Cnc, exposure times of $\sim$1 second are required to avoid detector saturation whilst for fainter targets (e.g. GJ 1214) the cadence of Twinkle observations would be of the order of several minutes.

\section{Methodology}

Here we take the Twinkle design as discussed in \cite{savini}, \cite{jason}, and \cite{wells}. Twinkle is currently entering a Phase B design review and thus the technical specifications may change. An updated analysis will be published when the design is finalised.

\subsection{Twinkle Field of Regard}

Twinkle’s field of regard, the region of sky in which it can be pointed, is a cone, centred on the anti-sun vector and extending to $\pm$40$^{\circ}$ from the ecliptic plane. This field may be extended to up to 60$^{\circ}$ for a limited number of targets. Twinkle's sky visibility is therefore similar to that of CHEOPS \cite{cessa} but is in contrast to TESS \cite{ricker}, JWST \cite{gardner} and ARIEL \cite{tinetti_ariel}, all of which have continuous coverage around the ecliptic poles and a partial visibility of the whole sky at lower latitudes. Here, we compare Twinkle's 40° field of regard to the celestial coordinates of planet-hosting stars to determine the number of known and predicted targets within Twinkle’s view.

\begin{figure}[h!]
  \centering
  \includegraphics[width=0.45\textwidth]{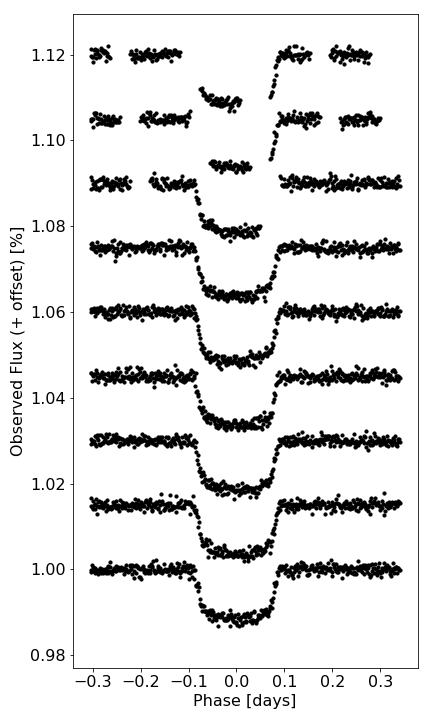}
  \includegraphics[width=0.45\textwidth]{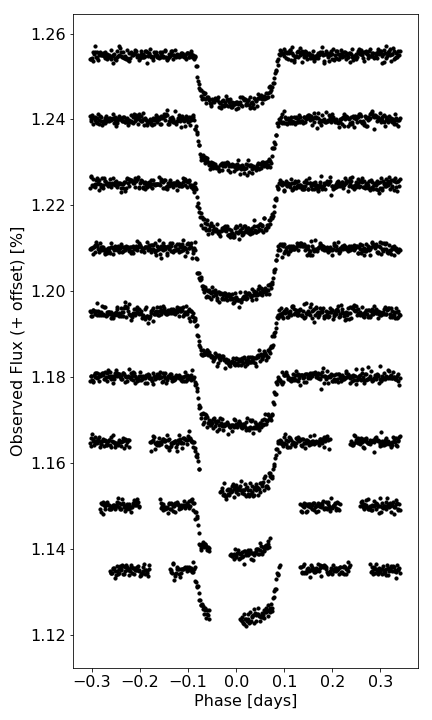}
  \caption{Modelled observations of partial and full transits of WASP-127 b in channel 1 at R = 250 for 2021. Of the 18 available transits in 2021, 12 of them can be observed by Twinkle in their entirety. The assumed stellar and planetary parameters are: RA = 160.56$^{\circ}$, Dec = -3.84$^{\circ}$, R$_*$ = 1.39 R$_\odot$, T$_*$ = 5650 K, Stellar K Mag = 8.64, R$_P$ = 15.03 R$_\oplus$, T$_P$ = 1533 K, Period = 4.18 days} \label{wasp127}
\end{figure}

Due to Twinkle’s low Earth orbit it will not always be possible to view the transits (or eclipses) in their entirety; instead, partial transits (or eclipses) will often be observed. For instance, Figure \ref{wasp127} shows modelled light curves for observations of all observable transits of WASP-127b during the year 2021; this includes 12 complete transits along with 6 partial transits. The radiometric model used here calculates the SNR achieved based upon observing a full transit or occultation. A follow-up paper will assess the impact of Earth obscuration and the subsequent effect on the number of potentially observable targets. It should be noted that exoplanet observations with Hubble are similarly restricted by Earth obscuration and that these partial transits will also occur for observations with CHEOPS.

\subsection{Observing techniques}

Twinkle will enable the study of exoplanets simultaneously at multiple wavelengths through transit, eclipse and phase-curve observations (see e.g. \cite{tinetti_spec} for an overview of the information content of these techniques for atmospheric characterisation). During transit, stellar light can be observed passing through the terminator region of the planet (transmission spectroscopy). Similarly, when the star eclipses the planet (i.e. the planet passes behind its host star in our line of sight) the flux difference resulting from the planet's dayside emission or reflection can be measured (emission or reflection spectroscopy). Phase curves are created by studying photometry of the star-planet system over a large portion of the planet’s orbit. Here, we consider transit and eclipse observations which, via spectroscopy, can allow for the atmospheric composition of the planet to be obtained. 

Additionally, photometric or low resolution observations in the optical and infrared can be utilised to refine planetary and orbital parameters, monitor stellar activity over time (e.g. \cite{bruno}) or search for transit time variations (TTVs, \cite{agol}) and transit duration variations (TDVs, \cite{kipping_exomoon}). TTVs have been extensively used as a valuable technique for finding additional planets within systems and constraining planetary masses and orbits (e.g. \cite{holman,lissauer,gillon,yee,linial}). Combined TTV and TDV signals are paramount to search for exomoons \cite{teachey}. The temporal binning required for Twinkle’s observations will depend upon the brightness of the host star but for brighter targets will be very short (<30 seconds) and therefore may be utilised to provide precise measurements of TTV and TDV due to low mass objects present in some planetary systems. For fainter stars, a temporal binning of a few minutes may be required, depending on the wavelength range considered. Whilst other future space facilities may also be able to detect TTV and TDV signals, Twinkle can obtain NIR light curves which exhibit highly reduced distortion from limb darkening and stellar activity. Additionally, multi-colour light curves significantly attenuate degeneracy of fitted limb darkening parameters across all wavelengths. Therefore Twinkle's capabilities for both high and low resolution transit (and eclipse) observations are considered here.

\subsection{Acquisition of Exoplanet Data and Missing Information}

Planetary data was downloaded from NASA’s Exoplanet Archive in order to account for all confirmed planets before being filtered such that only transiting planets were considered. The database was last accessed on 31st July 2018. However, the major exoplanet catalogues are sometimes incomplete and thus an effort has been made here to combine them (for a review of the current state of exoplanet catalogues see \cite{christiansen}). 

Hence, the data was verified, and in some cases gaps filled, utilising the Open Exoplanet Catalogue \cite{rein}, exoplanet.eu \cite{schneider} and TEPCat \cite{southworth}. If parameter values differed between sources, an average was taken. Planets not included in the NASA Exoplanet Archive were not added to the analysis to ensure that only confirmed planets were utilised.

\begin{table}[tbp]
\centering
\caption{Assumed molecular weight by planetary classification}
\label{tab:PlanetaryClass}
\begin{tabular}{lll}
\hline
Planet Class                 & Radius Bounds                                                          & Mean Molecular Weight (a.m.u.) \\ \hline
Earth                        & $R_p$ $\leq$ 1 R$_{\oplus}$                                            & 18                             \\
\multirow{2}{*}{Super-Earth} & 1 R$_{\oplus}$ \textless  $ R_p$ $\leq$ 1.7 R$_{\oplus}$ & 18                             \\
                             & 1.7 R$_{\oplus}$ \textless  $ R_p$ $\leq$ 2 R$_{\oplus}$ & 2.6                            \\
Sub-Neptune                  & 2 R$_{\oplus}$ \textless  $ R_p$ $\leq$ 3.5 R$_{\oplus}$ & 2.6                            \\
Neptune                      & 3.5 R$_{\oplus}$ \textless  $R_p$ $\leq$ 6 R$_{\oplus}$ & 2.6                            \\
Jupiter                      & $R_p$ \textgreater  6 R$_{\oplus}$                       & 2.3                            \\ \hline
\end{tabular}
\end{table}

Unknown parameters were inferred based on the following assumptions:

\begin{itemize}
   \item If the inclination is known, the impact parameter is calculated from:
    \begin{equation}
   b = \frac{ acos(i) }{R_*}
   \end{equation}
   \item Else, b = 0.5 (i.e. the midpoint of the equator and limb of the star)
   \item Planetary effective temperature (T$_p$) is estimated from: 
   \begin{equation}
   T_p =  T_* \left (\frac{ \sqrt[]{1- A} R_* }{2a\epsilon} \right )^{1/2}
   \end{equation}
   
   where a greenhouse effect of $\epsilon = 0.8 $ and a planetary albedo of $A =$ 0.3 (T$_P$ < 700K) or $ A =$ 0.1 (T$_p$ > 700K) are assumed \cite{tessenyi,seager}
   
   \item Planetary mass (M$_p$) was estimated utilising Forecaster \cite{chen}
   \item Atmospheric molecular mass was assumed from planetary class as stated in Table \ref{tab:PlanetaryClass}
   
\end{itemize}

\subsection{Twinkle Radiometric Model and Assessing Target Suitability}
The Twinkle Radiometric Model has been created using the methods described for calculating the signal and noise contributions for other exoplanet infrared spectroscopic missions in \cite{puig,sarkar}. The radiometric model simulates observational and instrumentation effects, utilising target characteristics to assess whether emission or transmission spectroscopy is preferable and to estimate the required number of observations to achieve a desired resolving power and signal to noise ratio (SNR).

The downloaded target information was inserted into the model and the average SNR across each spectrometer for one transit or eclipse was obtained for the maximum resolving power. The resolving power can be decreased from its native value in order to increase the SNR per wavelength bin. Stacking multiple observations increases the SNR of the final dataset and for this exercise the increase in SNR has been calculated from:

\begin{equation}
SNR_N = \sqrt[]{N} \times SNR_1
\end{equation}

where N is the number of observations.

\subsection{Future Planet Discoveries}
TESS and other surveys are predicted to discover thousands of planets around bright stars. In the first two years of operation, TESS is anticipated to detect over 4500 planets around bright stars and more than 10,000 giant planets around fainter stars \cite{barclay}. Here, these predicted TESS planets around brighter stars are incorporated into the analysis to highlight Twinkle’s capabilities to study future discoveries. The MAST archive has been utilised to obtain stellar parameters for the host stars of these predicted planets.

\subsection{Spectral Retrievals}

Determining the composition of exoplanetary atmospheres, and thus gaining an understanding of the atmospheric properties, provides insight about the processes occurring on these planets, as well as the presence or absence of clouds – an important constraint to understand atmospheric dynamics. Many molecules have absorption lines within Twinkle’s spectral bands including H$_2$O, carbon-bearing molecules (CO$_2$, CO, CH$_4$, C$_2$H$_2$, C$_2$H$_4$, C$_2$H$_6$), exotic metallic compounds (TiO, VO, SiO, TiH) and nitrogen/sulphur-bearing species (H$_2$S, SO$_2$, NH$_3$ and HCN). 

We have performed several spectral retrieval simulations to assess the information content of the spectra obtainable through Twinkle observations. The known population of exoplanets is diverse and thus to better assess Twinkle’s capabilities we focus here on three well known, but very different, planets: HD 209458 b, GJ 3470 b and 55 Cnc e. These planets all orbit very bright stars with K magnitudes of 6.31, 7.99 and 4.02 respectively.

To simulate emission and transmission forward models and atmospheric retrievals we use the open-source exoplanet atmospheric retrieval framework Tau-REx \cite{waldmann_2,waldmann_1} \url{https://github.com/ucl-exoplanets/TauREx_public}). We assume plane parallel atmospheres with 100 layers and include the contributions of collision-induced absorption (cia) of H$_2$-H$_2$ and H$_2$-He, Rayleigh scattering and grey-clouds. We use cross-section opacities calculated from the ExoMol database \cite{yurchenko} where available and from HITEMP \cite{rothman} and HITRAN \cite{gordon} otherwise. We have adopted Tau-REx to simulate retrievals at various resolutions. The assumed planetary characteristics are contained in Table \ref{tab:spec retrievals}. For HD 209458 b we take the atmospheric composition retrieved from observations with Hubble WFC3 \cite{tsiaras_hd209,macdonald}. We note that, whilst the spectral range for WFC3 is ideal for detecting water, constraining other molecules is more difficult with many models fitting the data. These degeneracies cause the retrieved abundances to differ from those expected from chemistry models \cite{venot_chem}.

\begin{table}[h!]
\centering
\caption{Planetary characteristics utilised for simulating atmospheric retrievals}
\label{tab:spec retrievals}
\resizebox{\textwidth}{!}{%
\begin{tabular}{lllllll}
\hline
Planet      & \begin{tabular}[c]{@{}l@{}}Equilibrium\\ Temperature {[}K{]}\end{tabular} & Clouds? & \begin{tabular}[c]{@{}l@{}}Mean Molecular\\ Weight\end{tabular} & \multicolumn{3}{l}{Molecular Abundances (Source)}                                                                                                                                 \\ \hline
55 Cnc e    & 2000                                                                      & No      & 2.33                                                            & \begin{tabular}[c]{@{}l@{}}CO\\ C$_2$H$_2$\\ HCN\end{tabular}            & \begin{tabular}[c]{@{}l@{}}1x10$^{-3}$\\ 1x10$^{-5}$\\ 1x10$^{-5}$\end{tabular}                   & \cite{tsiaras_55cnce}     \\ \hline
GJ 3470 b   & 700                                                                       & Yes     & 2.54                                                            & \begin{tabular}[c]{@{}l@{}}H$_2$0\\ CH4\\ CO\\ NH$_3$\\ CO$_2$\end{tabular} & \begin{tabular}[c]{@{}l@{}}1x10$^{-2}$\\ 4x10$^{-3}$\\ 1x10$^{-3}$\\ 1x10$^{-4}$\\ 1x10$^{-5}$\end{tabular} & \cite{venot_gj1214}      \\ \hline
HD 209458 b & 1000                                                                      & Yes     & 2.34                                                            & \begin{tabular}[c]{@{}l@{}}H$_2$0\\ HCN\\ NH$_3$ \\ CH$_4$\end{tabular}             & \begin{tabular}[c]{@{}l@{}}1x10$^{-5}$\\ 1x10$^{-6}$\\  1x10$^{-6}$\\ 1x10$^{-8}$\end{tabular}                   & \begin{tabular}[c]{@{}l@{}} \cite{tsiaras_hd209} \\ \cite{macdonald} \end{tabular} \\ \hline
\end{tabular}%
}
\end{table}

\section{Results}

\subsection{Targets within Twinkle's Field of Regard}

\begin{figure}[!ht]
  \centering
  \includegraphics[width=0.91\textwidth]{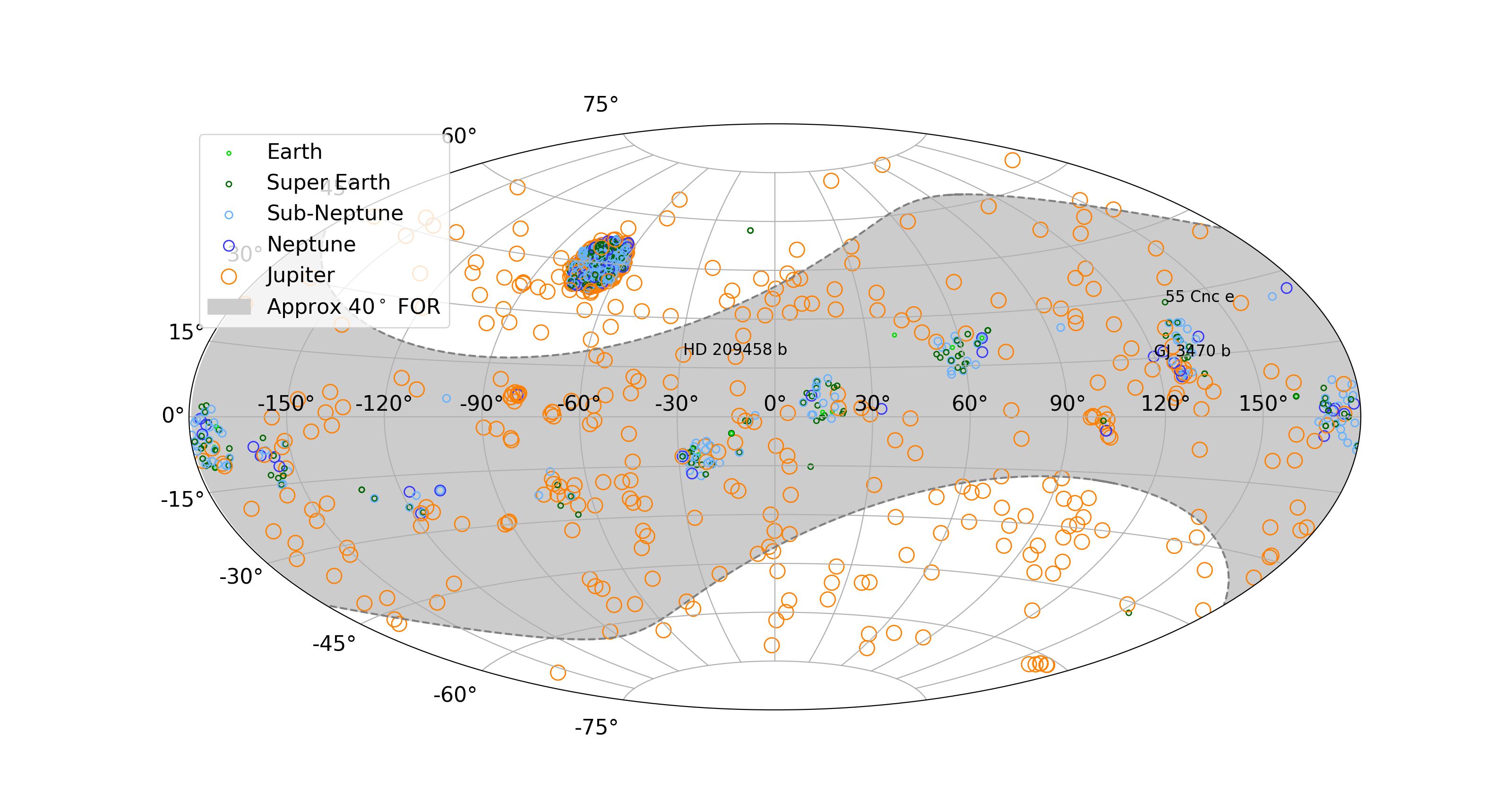}
  \includegraphics[width=0.91\textwidth]{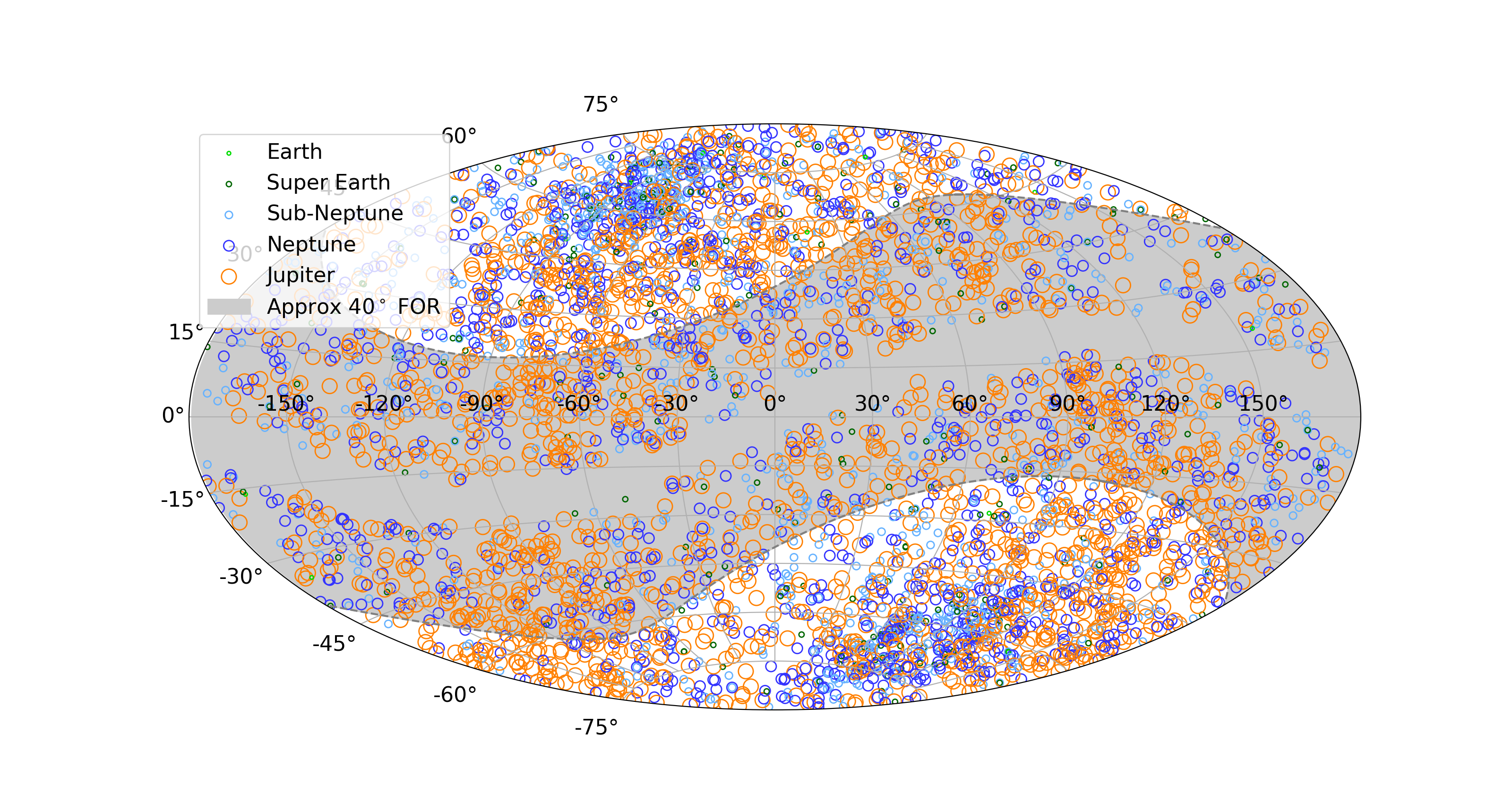}
  \includegraphics[width=0.91\textwidth]{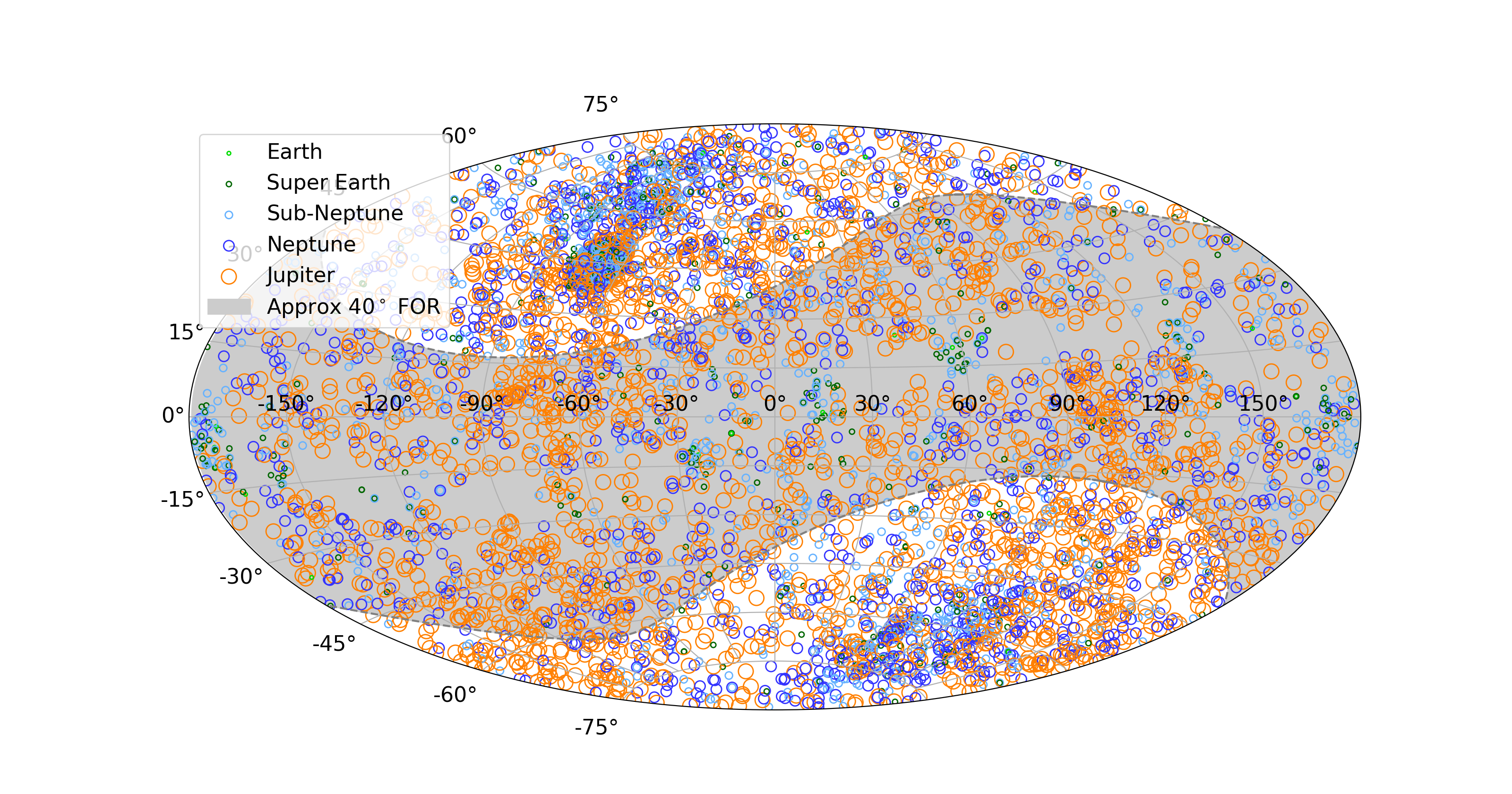}
  \caption{Top: Currently-known transiting exoplanets, their planetary classification and the sweep of Twinkle's field of regard. The initial Kepler field is the densely populated region at RA -70$^{\circ}$, Dec 45$^{\circ}$ and the locations of some well-known exoplanets are noted. Middle: Predicted TESS planet detections. Bottom: Currently-known planets and predicted TESS detections} \label{Planet Locations}
\end{figure}

NASA's Exoplanet Archive contains 2974 known transiting planets (as of July 2018) and Figure \ref{Planet Locations} displays the locations of these exoplanets and highlights those within the sweep of Twinkle’s field of regard. We note that the original Kepler field (which accounts for 2289 transiting planets) lies far from the ecliptic and thus cannot be observed with a 40$^{\circ}$ field of regard centred on the anti-sun vector. An extension of the field of regard to 60° from the ecliptic may allow some of these targets to be observed. However, Twinkle is best-suited for observing planets around bright stars and thus is not impeded by being unable to observe planets in the Kepler field which are generally hosted by fainter stars. Of the 685 transiting planets not discovered by the original Kepler mission, 548 lie within Twinkle's field of regard.

The predicted TESS planets are shown in Figure \ref{Planet Locations} and we find that of the 4376 targets plotted, 1815 lie within the 40$^{\circ}$ field of regard of Twinkle. TESS is also predicted to discover more than 10,000 planets around fainter stars and whilst some of these will be within Twinkle’s field of regard, many are unlikely to be suitable targets for observation with Twinkle. However, some larger, hotter planets may be observable.

We require that all targets in our study must, as a minimum, have the following known parameters:

\begin{itemize}
 	\item Stellar data:
    \begin{itemize}
	\item Right ascension and declination
	\item Mass
	\item Radius
	\item K Mag
	\item 
    
     3070 K < $T_*$ < 7200 K
    
    \end{itemize}
    
	\item Planet data:
    \begin{itemize}
	\item Radius
	\item Period
    \end{itemize}
    
\end{itemize}

\begin{figure}[!ht]
  \centering
  \includegraphics[width=1\textwidth]{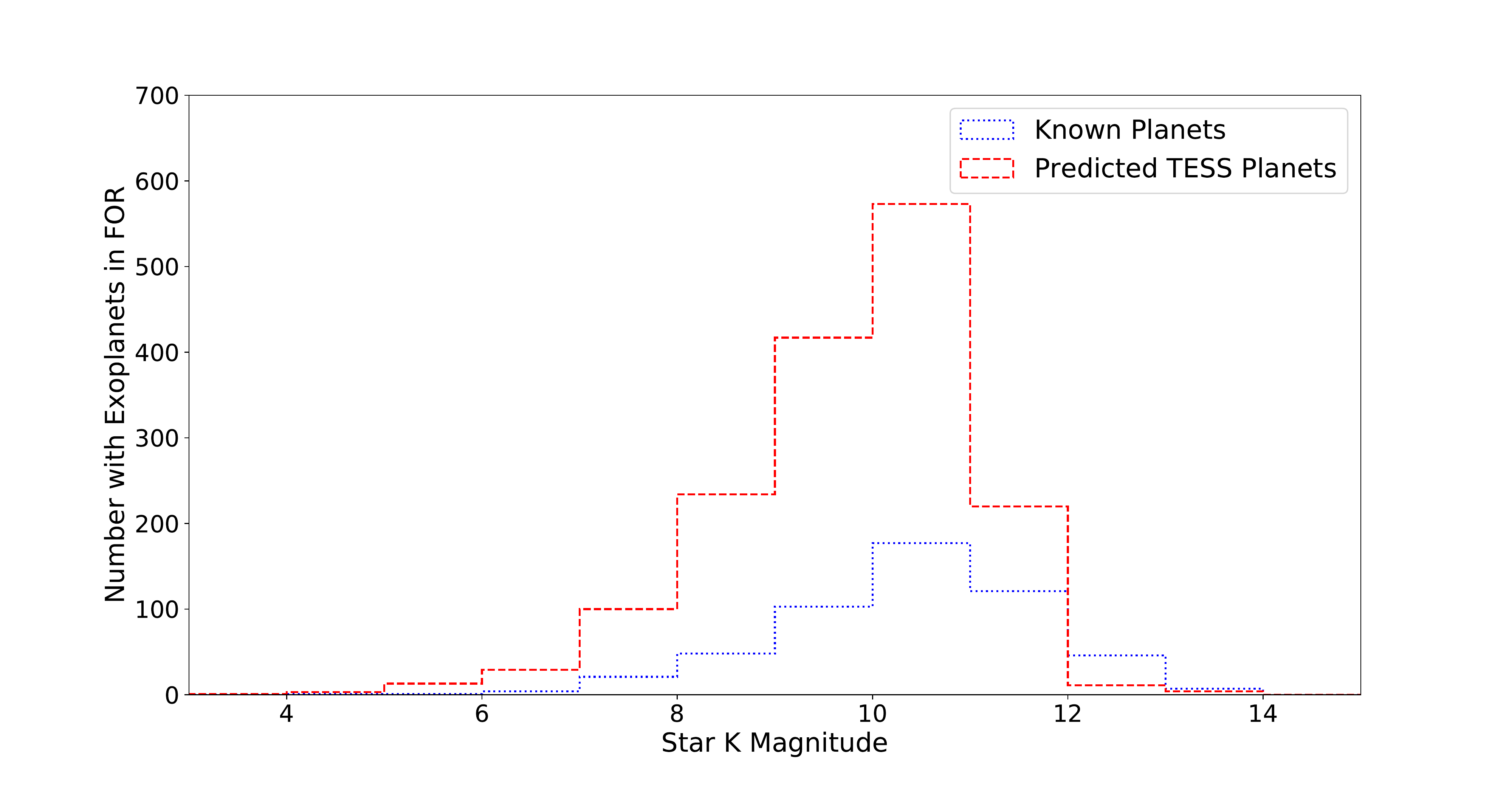}
  \caption{Stellar brightness in K mag for known planets and predicted TESS detections within Twinkle’s field of regard} \label{starmag}
\end{figure}

Of the 548 currently known planets that Twinkle could observe, 530 were sufficiently characterised for inclusion in this analysis. For the simulated TESS planets within Twinkle’s field of regard, 1608 could be analysed. Figure \ref{starmag} shows stellar magnitudes that could currently be probed by Twinkle. Planets are found to be most commonly orbiting stars of with K magnitudes between 9 and 12. The magnitudes of the simulated TESS target stars are also plotted here, and it is clear that many more planets are expected to be found around bright stars within Twinkle’s field of regard.

\subsection{Potentially Observable Targets}
\begin{figure}[!ht]
  \centering
  \includegraphics[width=0.7\textwidth]{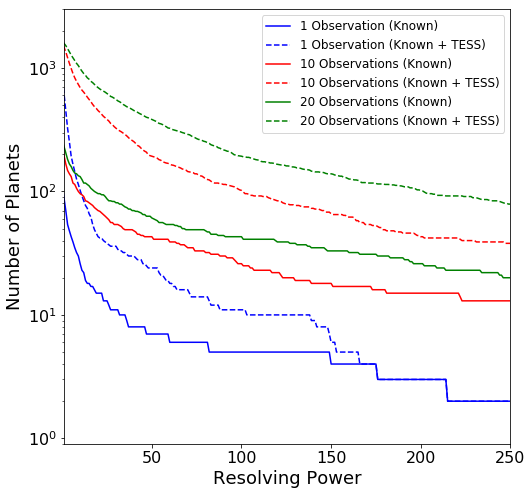}
  \caption{Number of known and predicted TESS planets with Twinkle’s field of regard for which SNR $\geq$ 7 is achievable  at a given resolving power in channel 0 (0.4 - 1.0$\mu$m)} \label{achievable_res_ch0}
\end{figure}

\begin{figure}[ht!]
  \centering
  \includegraphics[width=0.7\textwidth]{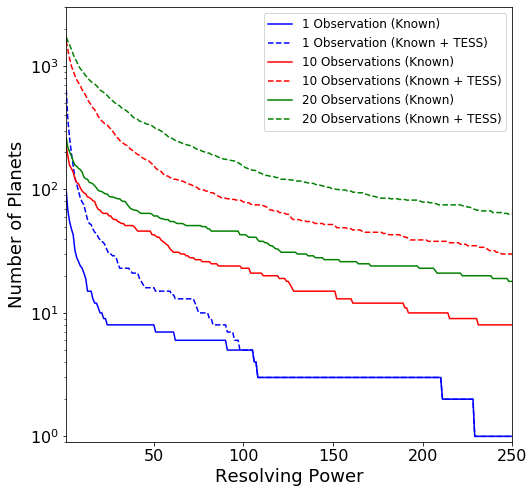}
  \caption{Number of known and predicted TESS planets with Twinkle’s field of regard for which SNR $\geq$ 7 is achievable  at a given resolving power in channel 1 (1.3 - 2.42$\mu$m)} \label{achievable_res_ch1}
\end{figure}

\begin{figure}[ht!]
  \centering
  \includegraphics[width=0.7\textwidth]{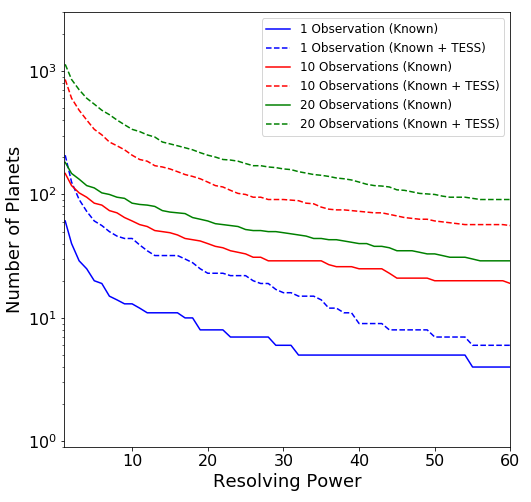}
  \caption{Number of known and predicted TESS planets with Twinkle’s field of regard for which SNR $\geq$ 7 is achievable  at a given resolving power in channel 2 (2.42 - 4.5$\mu$m)} \label{achievable_res_ch2}
\end{figure}

\begin{figure}[p]

  \includegraphics[width=14cm]{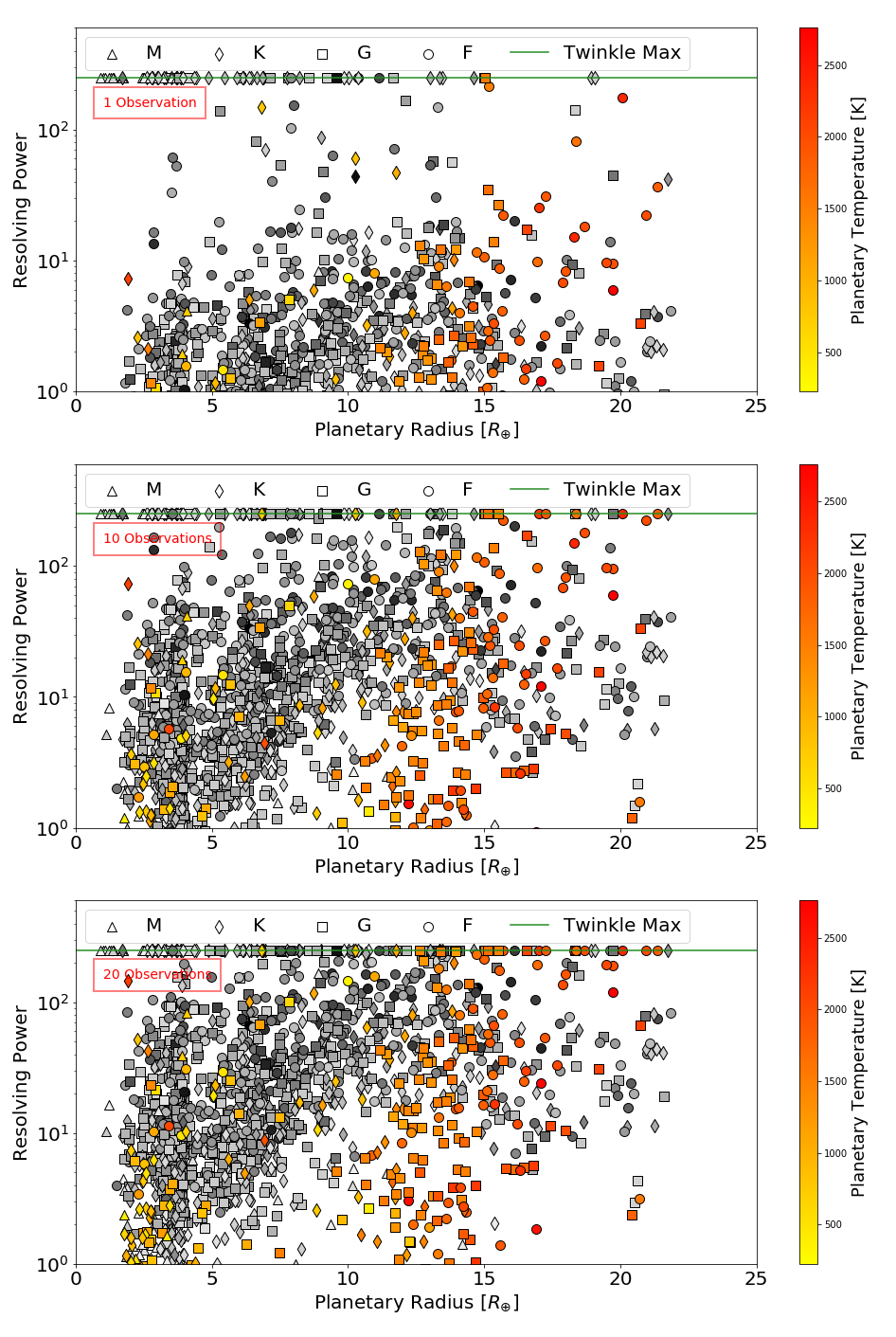}
  \caption{Achievable resolving power in infrared channel 0 (0.4 - 1.0$\mu$m) for currently known (coloured) and expected TESS planets (grey) assuming a requirement of SNR $\geq$ 7 for a given number of transit or eclipse observations. The shape of the points indicates the stellar type of the host star and the green line represents Twinkle's maximum resolving power in the channel} \label{tess_res_ch0}
\end{figure}

\begin{figure}[p]

  \includegraphics[width=14cm]{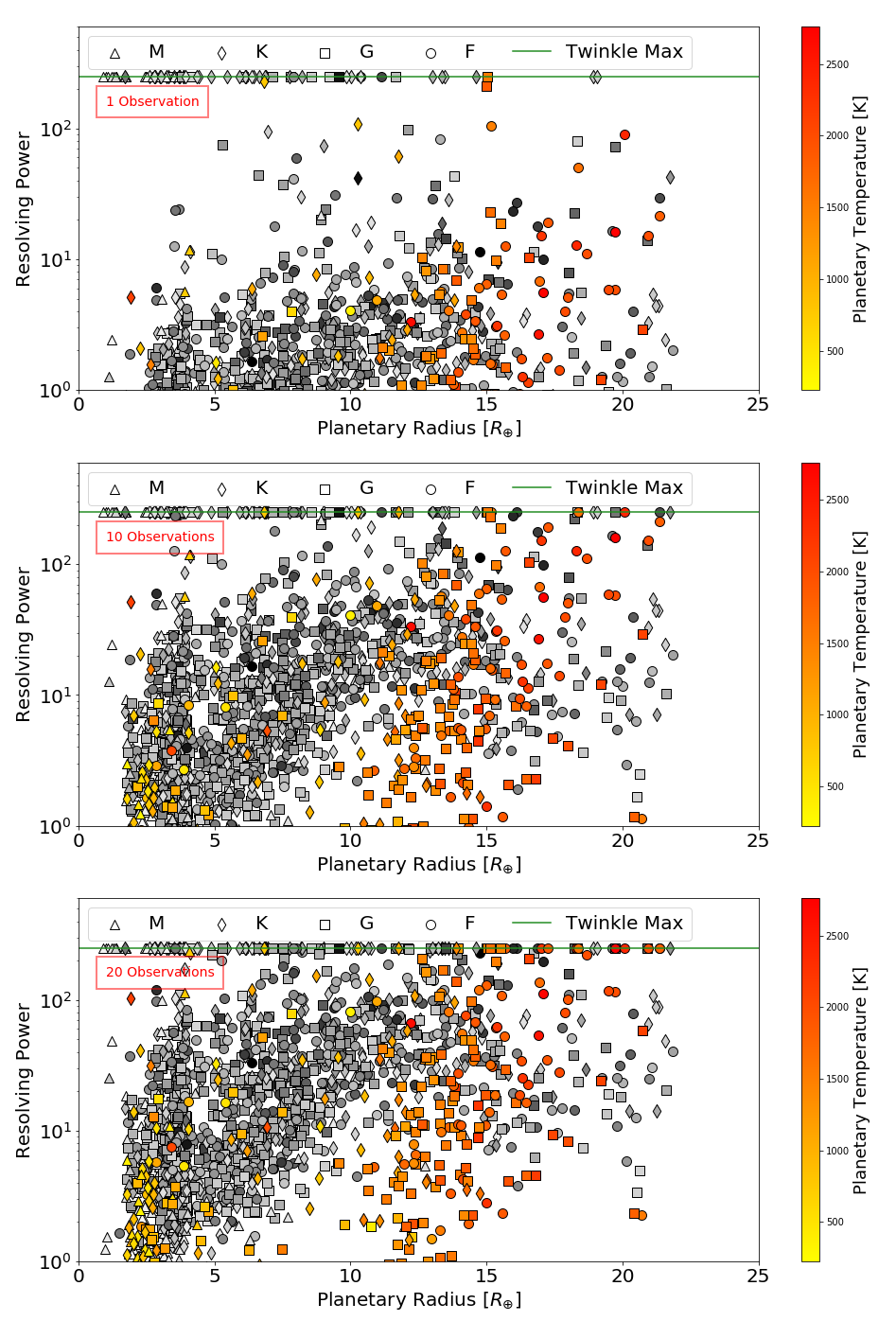}
  \caption{Achievable resolving power in infrared channel 1 (1.3 - 2.42$\mu$m) for currently known (coloured) and expected TESS planets (grey) assuming a requirement of SNR $\geq$ 7 for a given number of transit or eclipse observations. The shape of the points indicates the stellar type of the host star and the green line represents Twinkle's maximum resolving power in the channel} \label{tess_res_ch1}
\end{figure}

\begin{figure}[p]
  
  \includegraphics[width=14cm]{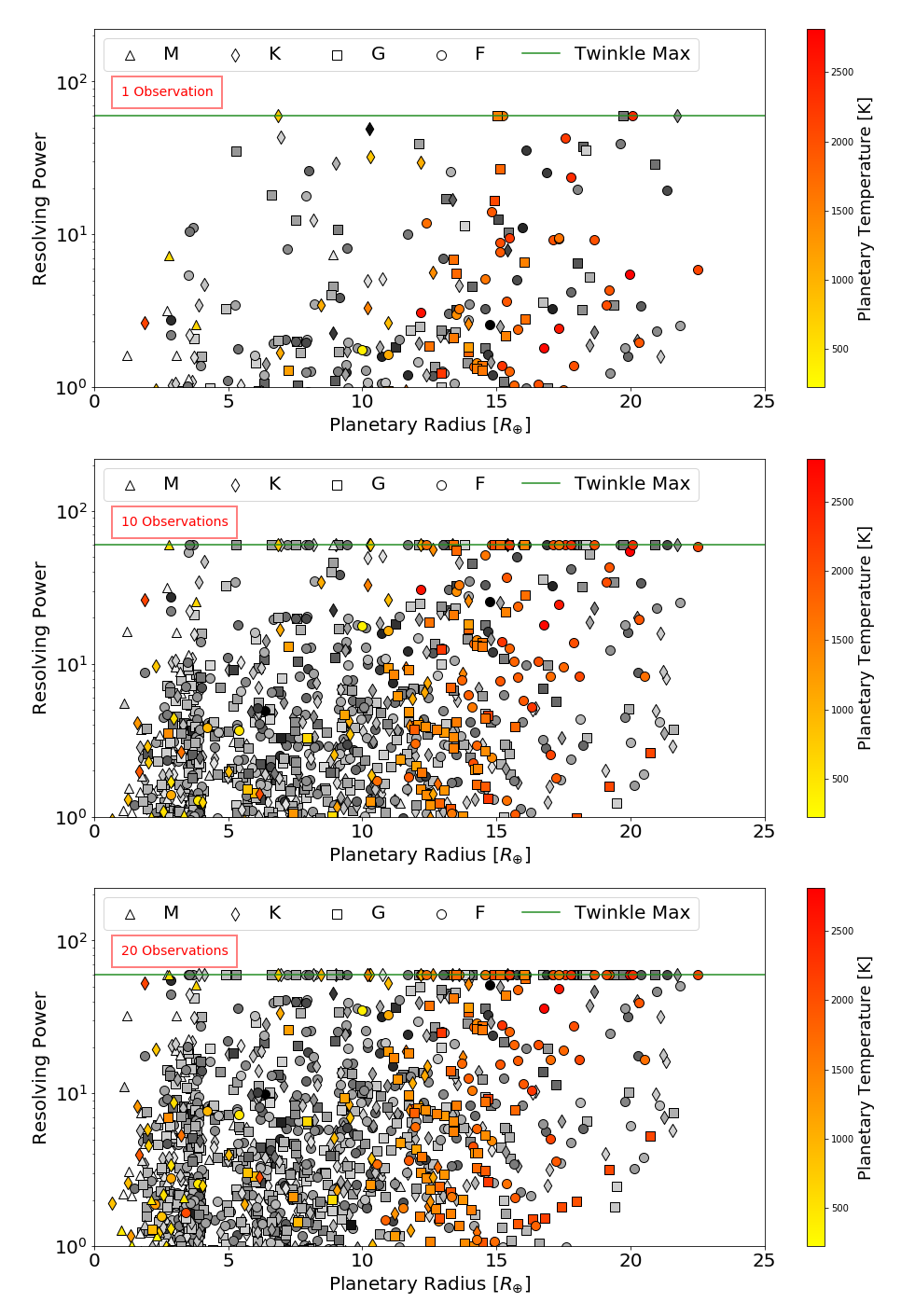}
  \caption{Achievable resolving power in infrared channel 2 (2.42 - 4.5$\mu$m) for currently known (coloured) and expected TESS planets (grey) assuming a requirement of SNR $\geq$ 7 for a given number of transit or eclipse observations. The shape of the points indicates the stellar type of the host star and the green line represents Twinkle's maximum resolving power in the channel} \label{tess_res_ch2}
\end{figure}

Defining a requirement for SNR $\geq$ 7,  Figures \ref{achievable_res_ch0}, \ref{achievable_res_ch1} and \ref{achievable_res_ch2} display the number of planets that could be observed in channels 0, 1 and 2, at a given resolution, for a given number of transit or eclipse observations. Assuming the same SNR requirement, the resolution achieved over a given number of transits or eclipses is displayed in Figures \ref{tess_res_ch0}, \ref{tess_res_ch1} and \ref{tess_res_ch2} showing the distribution of the radius of observable planets along with the planet’s temperature class. The SNR requirement will depend upon the user’s preferences and the observations undertaken and therefore should be taken only as an indicative value.

We find that in one observation Twinkle could study 89 known planets and 469 predicted TESS planets in channel 1 with R < 20 and 12 known planets and 29 predicted TESS planets at higher resolutions. With ten transits or eclipses, spectroscopy is possible for 81 known and 307 TESS planets at R\textgreater20, and for 144 known and 1041 TESS planets at lower resolution. In each case the majority of targets are large gaseous planets (e.g. 84.5\% Jupiters and 15\% Neptunes/Sub-Neptunes for R \textgreater 20 in 10 observations). With a larger number of observations, the atmospheres of 46 smaller, potentially rocky planets (R\textless1.7R$_\oplus$) are characterisable using low resolution spectroscopy (R\textless20) in channel 1 with 20 observations. The relation between stellar magnitude, planetary radius and the achievable resolving power is shown in Appendix A.

\clearpage

\subsection{Spectral Retrievals}
For the three chosen planets, we calculate the resolution achievable to reach an SNR > 7 for each channel with a given number of transits/eclipses and run Tau-REx spectral retrievals with the observation parameters contained in Table \ref{tab:retrieval parameters}.

\begin{table}[ht!]
\centering
\caption{Parameters of the simulated atmospheric retrievals with Tau-REx}
\label{tab:retrieval parameters}
\resizebox{\textwidth}{!}{%
\begin{tabular}{lllll}
\hline
Planet      & Number of Observations & R ($\lambda$ \textless 2.42$\mu$m) & R ($\lambda$ \textgreater 2.42$\mu$m) & Observation Type \\ \hline
HD 209458 b & 1                      & 250                   & 60                       & Transit          \\ 
GJ 3470 b   & 10                     & 65                    & 20                       & Transit          \\ 
55 Cnc e    & 10                     & 10                    & 20                       & Eclipse          \\ \hline
\end{tabular}%
}
\end{table}

\begin{figure}[ht!]
  \centering
  \includegraphics[width=1\textwidth]{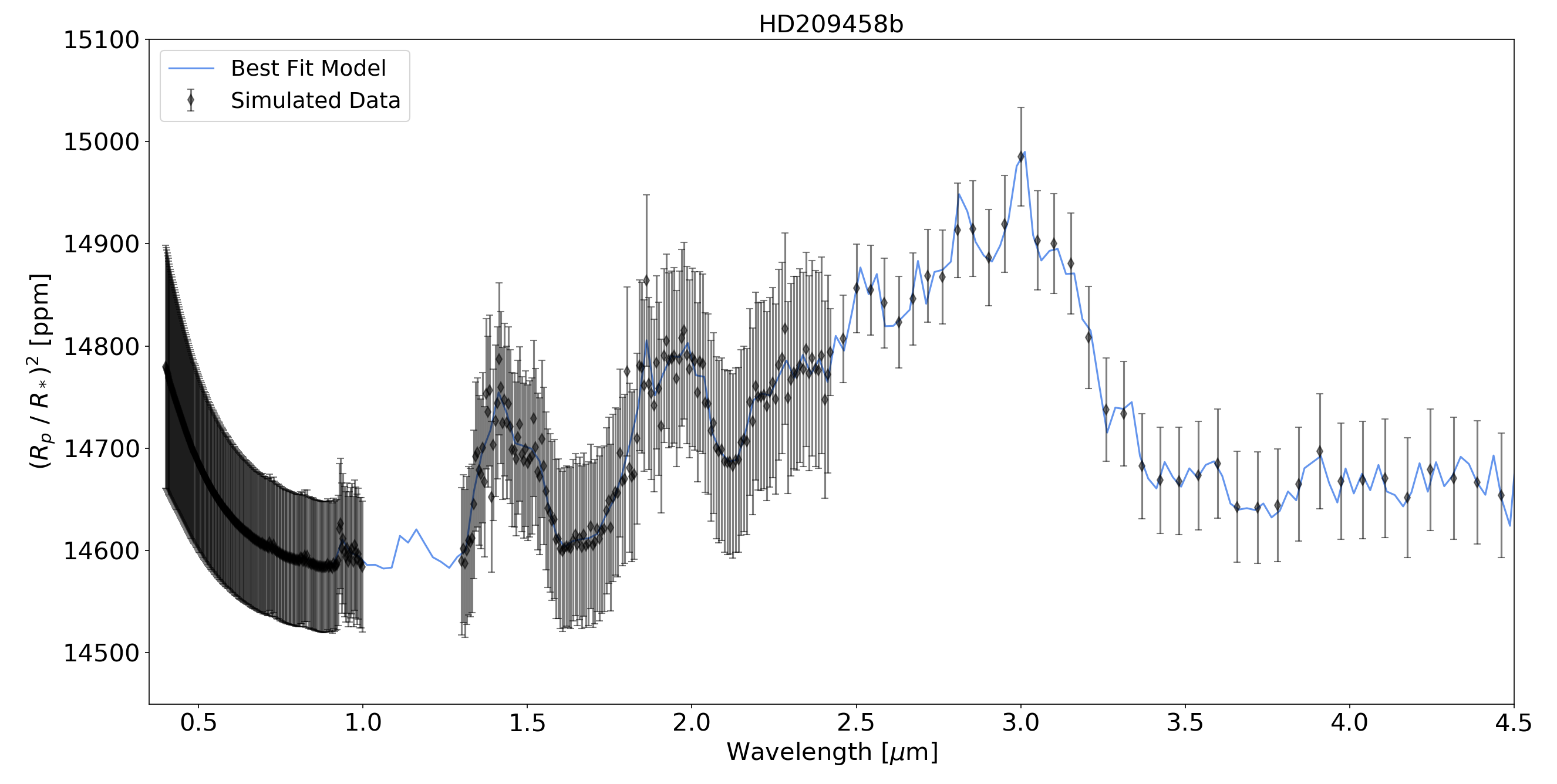}
  \includegraphics[width=1\textwidth]{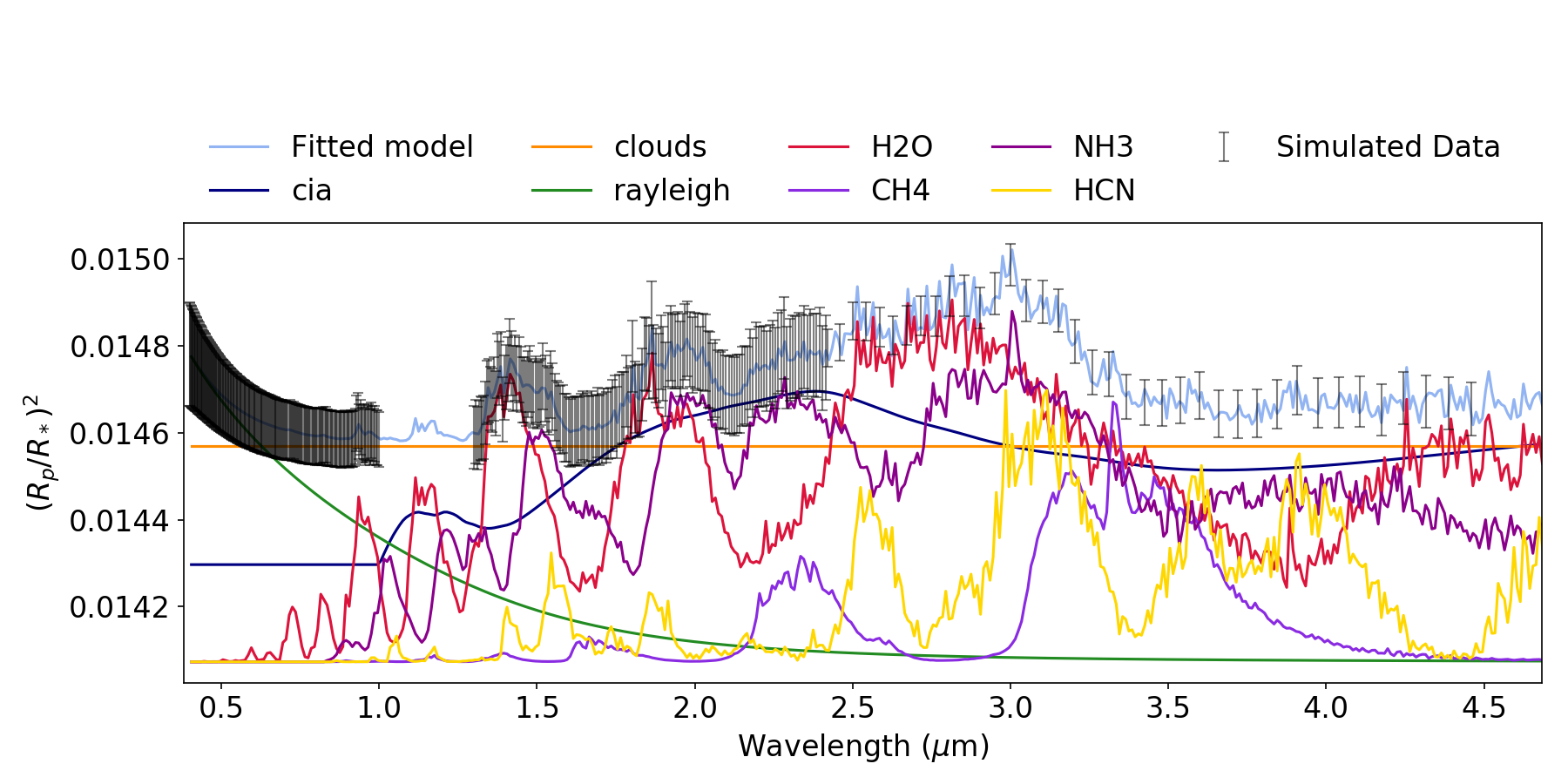}
  \caption{Spectral retrieval of HD 209458 b (input values: P$_{Clouds}$ = 1x10$^{4}$ Pa, H$_2$0 = 1x10$^{-5}$, HCN = 1x10$^{-6}$, NH$_3$ = 1x10$^{-6}$, CH$_4$ = 1x10$^{-8}$) at R = 250 ($\lambda$ < 2.42$\mu$m) and R = 60 ($\lambda$ > 2.42$\mu$m) with 1 complete transit observation which recovers the main molecular composition but does not constrain CH$_4$ due to the very low abundance of the molecule. The bottom panel shows the individual contributions of each molecule and of Rayleigh scattering, collision induced absorption (cia) and clouds. Note that, as shown in Figure \ref{wasp127}, more than one visit may be required to observe a full transit light curve and that this is the raw performance of the instrument and the plot may not show an optimised setting} \label{hd209spec}
\end{figure}

\begin{figure}[ht!]
  \centering
  \includegraphics[width=1\textwidth]{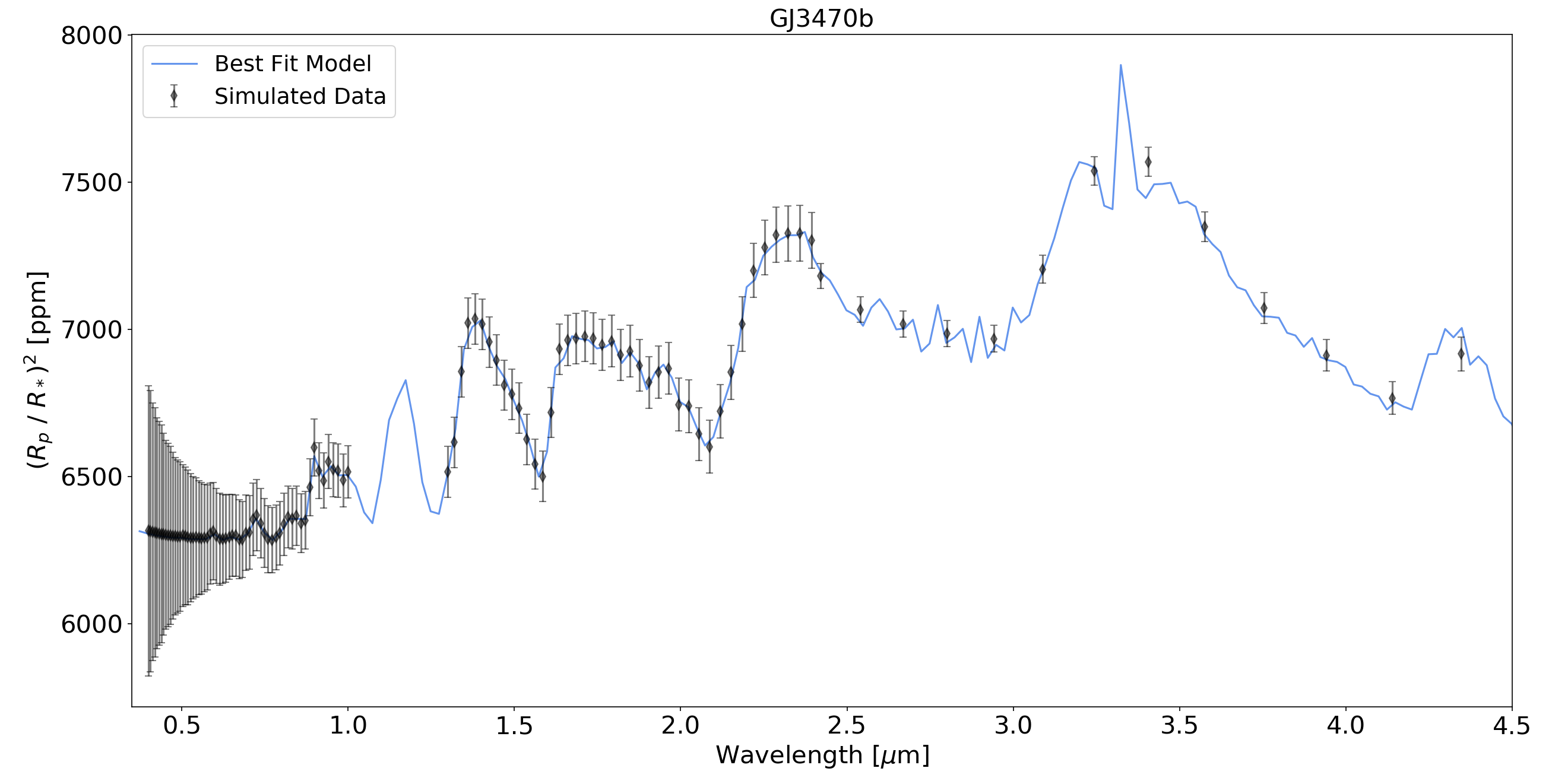}
  \includegraphics[width=1\textwidth]{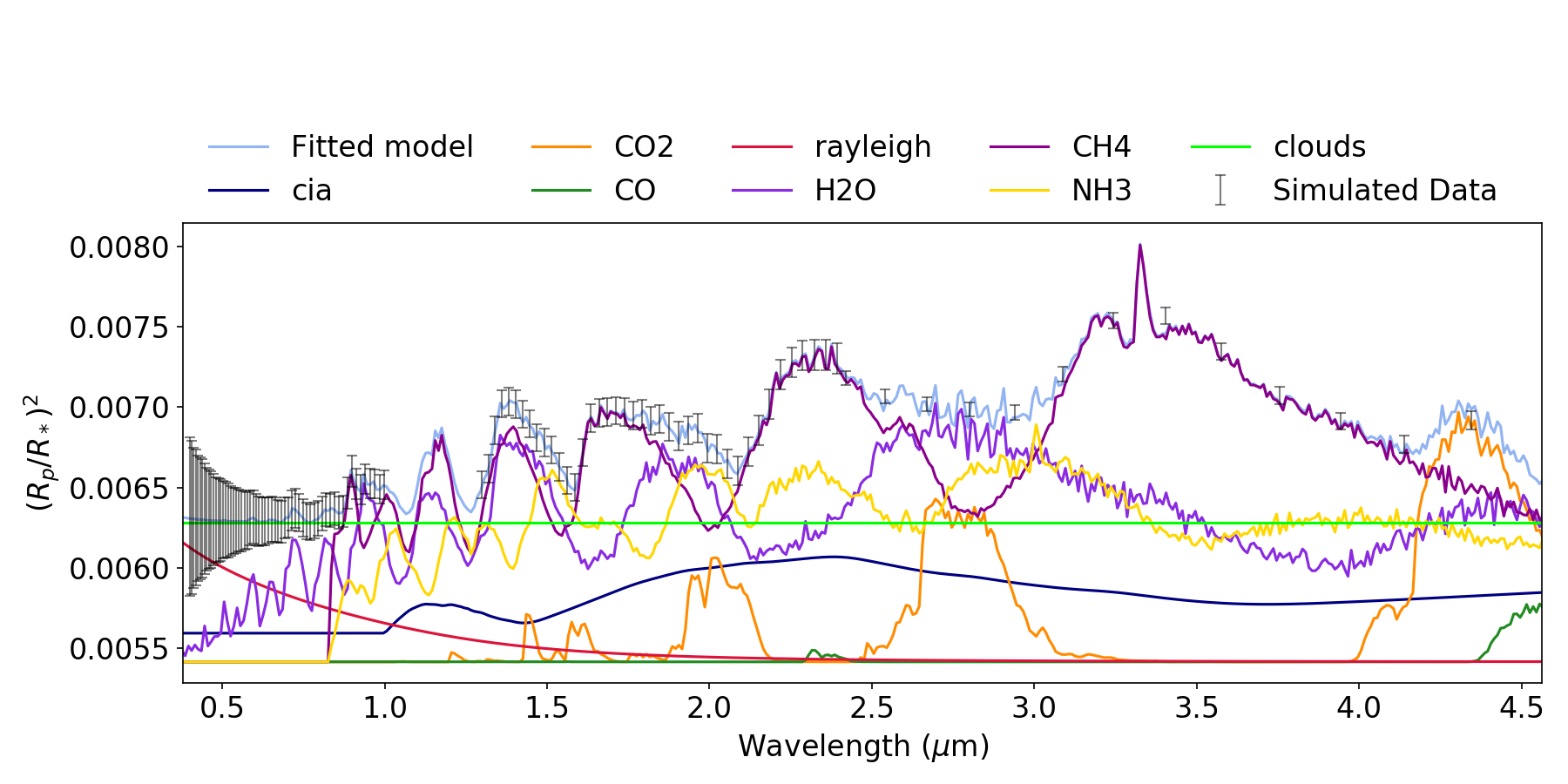}
  \caption{Spectral retrieval for GJ 3470 b (input values: P$_{Clouds}$ = 1x10$^{3}$ Pa, H$_2$0 = 1x10$^{-2}$, CH$_4$ = 4x10$^{-3}$, CO = 1x10$^{-3}$, NH$_3$ = 1x10$^{-4}$, CO$_2$ = 1x10$^{-5}$) at R = 65 ($\lambda$ < 2.42$\mu$m) and R = 20 ($\lambda$ > 2.42$\mu$m) with 10 complete transit observations which correctly recovers the major molecular abundances and cloud pressure but does not detect CO. The bottom panel shows the individual contributions of each molecule and of Rayleigh scattering, collision induced absorption (cia) and clouds. Note that, as shown in Figure \ref{wasp127}, more than ten visits may be required to observe ten full transit light curves} \label{gj3470spec}
\end{figure}

HD 209458 b could be observed, in a single transit, at the highest resolution possible with Twinkle’s instrumentation and the subsequent retrieval correctly identifies the abundances and cloud pressure as shown in Figure \ref{hd209spec} and Table \ref{tab:retrieval abundances}. Figure \ref{gj3470spec} displays the retrieval for GJ 3470 b and it can be seen that after 10 transits, at a resolution in channel 1 similar to the max resolution of Hubble WFC3, the molecular abundance of H$_2$O, CH$_4$, NH$_3$ and CO$_2$ have been correctly recovered, even with a cloudy atmosphere (assuming a grey cloud with a cloud top pressure of 10 mbar). However, as shown in Table \ref{tab:retrieval abundances} the CO abundance is not recovered as the strongest CO band accessible to Twinkle is at 4.7$\mu$m which is at the edge of its observable wavelength range and oftentimes masked by other molecular constituents.

\begin{figure}[ht!]
  \centering
  \includegraphics[width=1\textwidth]{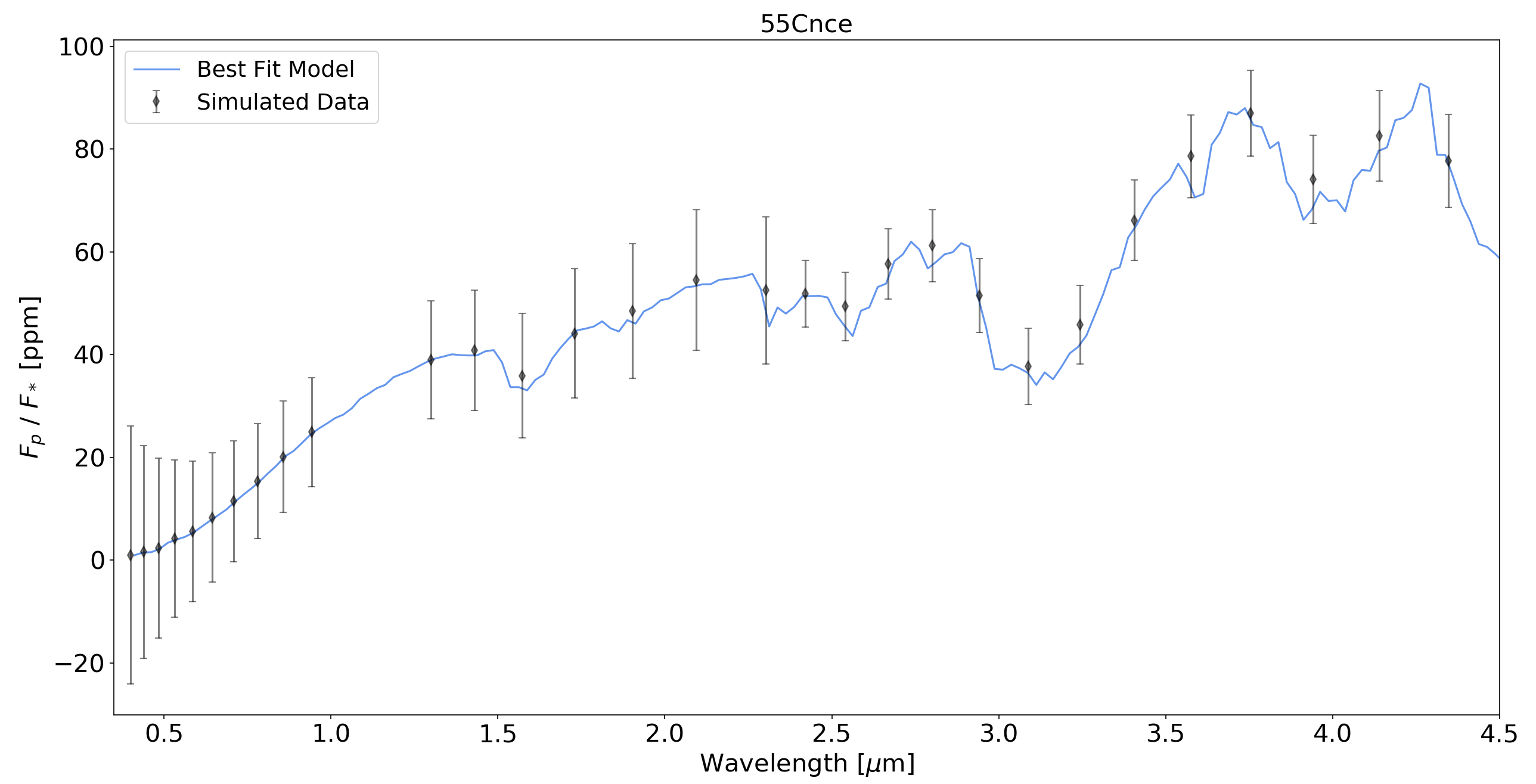}
  \includegraphics[width=1\textwidth]{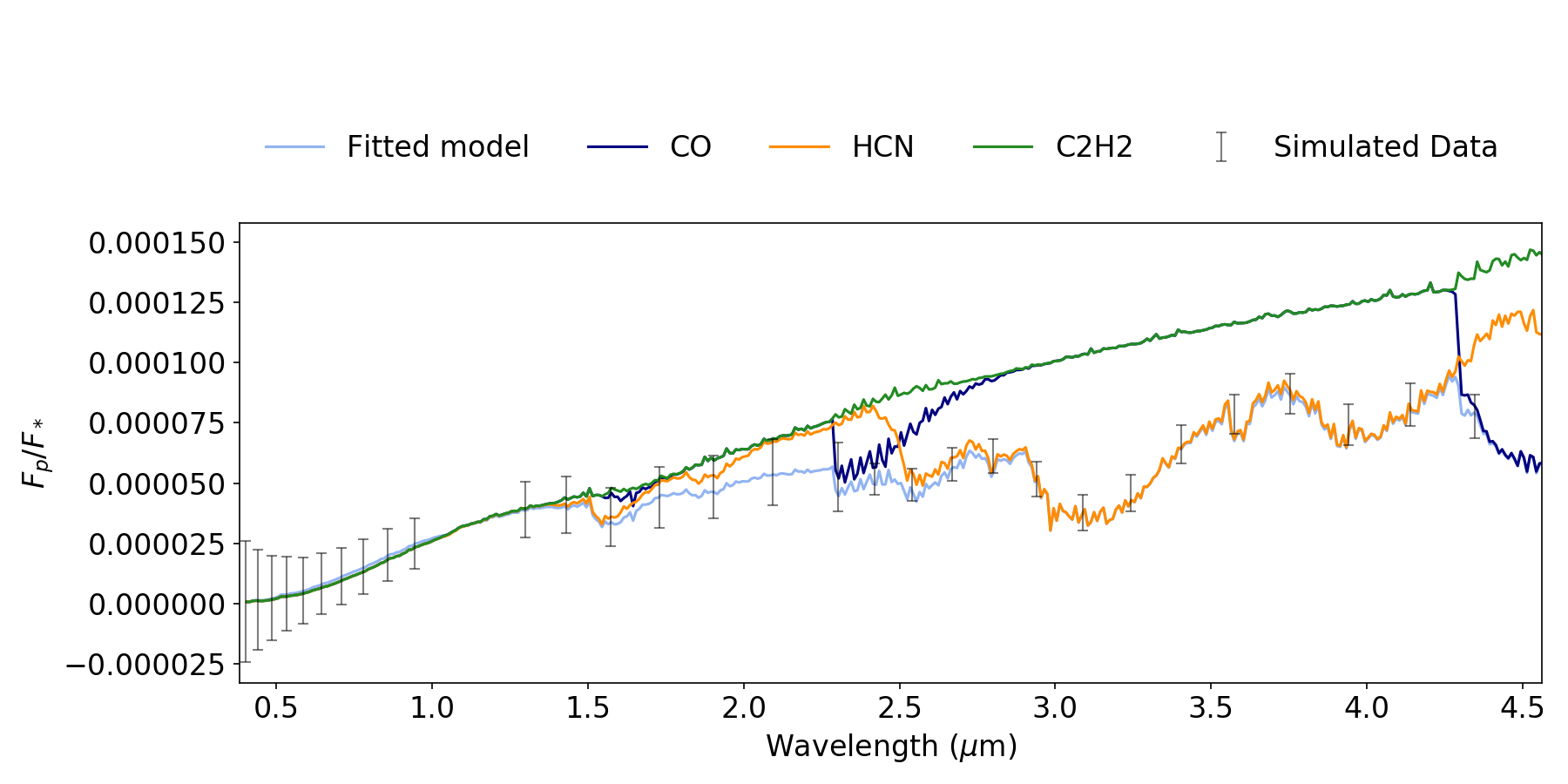}
  \caption{Spectral retrieval for 55 Cnc e (input values: cloud free, CO = 1x10$^{-3}$ Pa, C$_2$H$_2$ = 1x10$^{-5}$, HCN = 1x10$^{-5}$) at R = 10 ($\lambda$ < 2.42$\mu$m) and R = 20 ($\lambda$ > 2.42$\mu$m) with 10 complete eclipse observations which correctly recovers the HCN and CO abundances but does not constrain the C$_2$H$_2$ abundance. The bottom panel shows the individual contributions of each molecule and of Rayleigh scattering, collision induced absorption (cia) and clouds. Note that, as shown in Figure \ref{wasp127}, more than ten visits may be required to observe ten full eclipse light curves} \label{55cncespec}
\end{figure}

Due to its small size, 55 Cnc e is challenging to observe and thus 10 eclipse observations are required to obtain low resolution spectra. Despite this difficulty, Tau-REx retrievals of 55 Cnc e resulted in the main constituents of the simulated atmosphere (CO and HCN) being identified well (Figure \ref{55cncespec} and Table \ref{tab:retrieval abundances}). Increasing the number of visits to 20 results in the accurate recovery of the simulated abundances of CO and HCN as well as more accurately retrieving C$_2$H$_2$. The retrievability of C$_2$H$_2$ highlights the benefit of higher resolution spectra when measuring abundances of complex carbon molecules. The posterior plot for each of these retrievals is contained Appendix B.

\begin{table}[ht!]
\caption{The original and retrieved abundances for each planet from the simulated retrievals with Tau-REx. We note that we correctly retrieve the abundances for HD 209458 b as well as most of the abundances for GJ 3470 b and 55 Cnc e. However, the CO abundance is not recovered for GJ 3470 b and for 55 Cnc e C$_2$H$_2$ is also not constrained. This is due to other molecules obscuring the absorption features as can be seen in the contributions plots in Figures \ref{gj3470spec} and \ref{55cncespec}}
\label{tab:retrieval abundances}
\resizebox{\textwidth}{!}{%
\begin{tabular}{lllll}
\hline
Planet                       & Molecule              & Input Abundance (log10) & Retrieved Abundance (log10) & Retrieval Abundance Uncertainty (log10) \\ \hline
\multirow{8}{*}{HD 209458 b} & \multirow{2}{*}{H$_2$0}  & \multirow{2}{*}{-5.00}  & \multirow{2}{*}{-5.02}      & +0.17                                   \\
                             &                       &                         &                             & -0.17                                   \\
                             & \multirow{2}{*}{HCN}   & \multirow{2}{*}{-6.00}  & \multirow{2}{*}{-6.09}      & +0.33                                   \\
                             &                       &                         &                             & -0.44                                   \\
                             & \multirow{2}{*}{NH$_3$}  & \multirow{2}{*}{-6.00}  & \multirow{2}{*}{-5.98}      & +0.14                                   \\
                             &                       &                         &                             & -0.14                                   \\ & \multirow{2}{*}{CH$_4$}  & \multirow{2}{*}{-8.00}  & \multirow{2}{*}{-9.61}      & +1.54                                   \\
                             &                       &                         &                             & -1.63                                   \\ \hline
\multirow{10}{*}{GJ 3460 b}  & \multirow{2}{*}{H$_2$0}  & \multirow{2}{*}{-2.00}  & \multirow{2}{*}{-2.03}      & +0.36                                   \\
                             &                       &                         &                             & -0.51                                   \\
                             & \multirow{2}{*}{CH$_4$}  & \multirow{2}{*}{-2.40}  & \multirow{2}{*}{-2.45}      & +0.37                                   \\
                             &                       &                         &                             & -0.46                                   \\
                             & \multirow{2}{*}{CO}   & \multirow{2}{*}{-3.00}  & \multirow{2}{*}{-7.49}      & +3.39                                   \\
                             &                       &                         &                             & -3.01                                   \\
                             & \multirow{2}{*}{NH$_3$}  & \multirow{2}{*}{-4.00}  & \multirow{2}{*}{-4.15}      & +0.38                                   \\
                             &                       &                         &                             & -0.44                                   \\
                             & \multirow{2}{*}{CO$_2$}  & \multirow{2}{*}{-5.00}  & \multirow{2}{*}{-5.13}      & +0.44                                   \\
                             &                       &                         &                             & -0.52                                   \\ \hline
\multirow{6}{*}{55 Cnc e}    & \multirow{2}{*}{CO}   & \multirow{2}{*}{-3.00}  & \multirow{2}{*}{-3.10}      & +0.42                                   \\
                             &                       &                         &                             & -0.41                                   \\
                             & \multirow{2}{*}{HCN}  & \multirow{2}{*}{-5.00}  & \multirow{2}{*}{-5.02}      & +0.17                                   \\
                             &                       &                         &                             & -0.17                                   \\
                             & \multirow{2}{*}{C$_2$H$_2$} & \multirow{2}{*}{-5.00}  & \multirow{2}{*}{-7.47}      & +2.60                                   \\
                             &                       &                         &                             & -2.98                                   \\ \hline
\end{tabular}%
}
\end{table}

\newpage

\clearpage

\section{Discussion}

This first iteration of assessing Twinkle’s performance for exoplanetary science has shown that many planets are potentially observable with Twinkle. Twinkle is currently entering a Phase B design review and thus the technical specifications may change. An updated analysis will be published when the design is finalised.

We find that a large number of targets could be studied with multi-band photometry (i.e. photons binned into a limited number of broad photometric bands) or low resolution spectroscopy in a single observation. Simultaneous photometric measurements in the optical and infrared would allow for rigorous constraints on the planetary, stellar and orbital parameters of a system. Eclipse photometry in the visible and infrared provide the bulk temperature and albedo of the planet, thereby allowing an estimate of the planetary energy balance and insight into whether the planet has an additional energy input, such as an internal heat source. For the planets discovered by TESS and other transit surveys, constraining planetary ephemerides will be a key requirement to allow for spectroscopic observations further into the future. The planets which have deeper transits could be observed from the ground but fainter and shallower transits will require space-based facilities. Twinkle may also be able to detect TTVs and TDVs but this capability will depend upon the cadence of the observations as well as the impact of Earth obscuration and therefore requires further analysis on specific targets. The temporal binning for brighter targets will be very short (\textless 30 seconds) whilst for fainter stars, a temporal binning of a few minutes may be required, depending on the wavelength range considered. Such a cadence is suitable for TTV and TDV analysis and Twinkle's ability to obtain IR light curves will be useful for studying limb darkening and stellar activity which are expected to exhibit highly reduced distortion over this spectral range compared to visible wavelengths. Additionally, multi-colour light curves significantly attenuate degeneracy of fitted limb darkening parameters across all wavelengths.

\begin{figure}[ht!]
  \centering
  \includegraphics[width=0.85\textwidth]{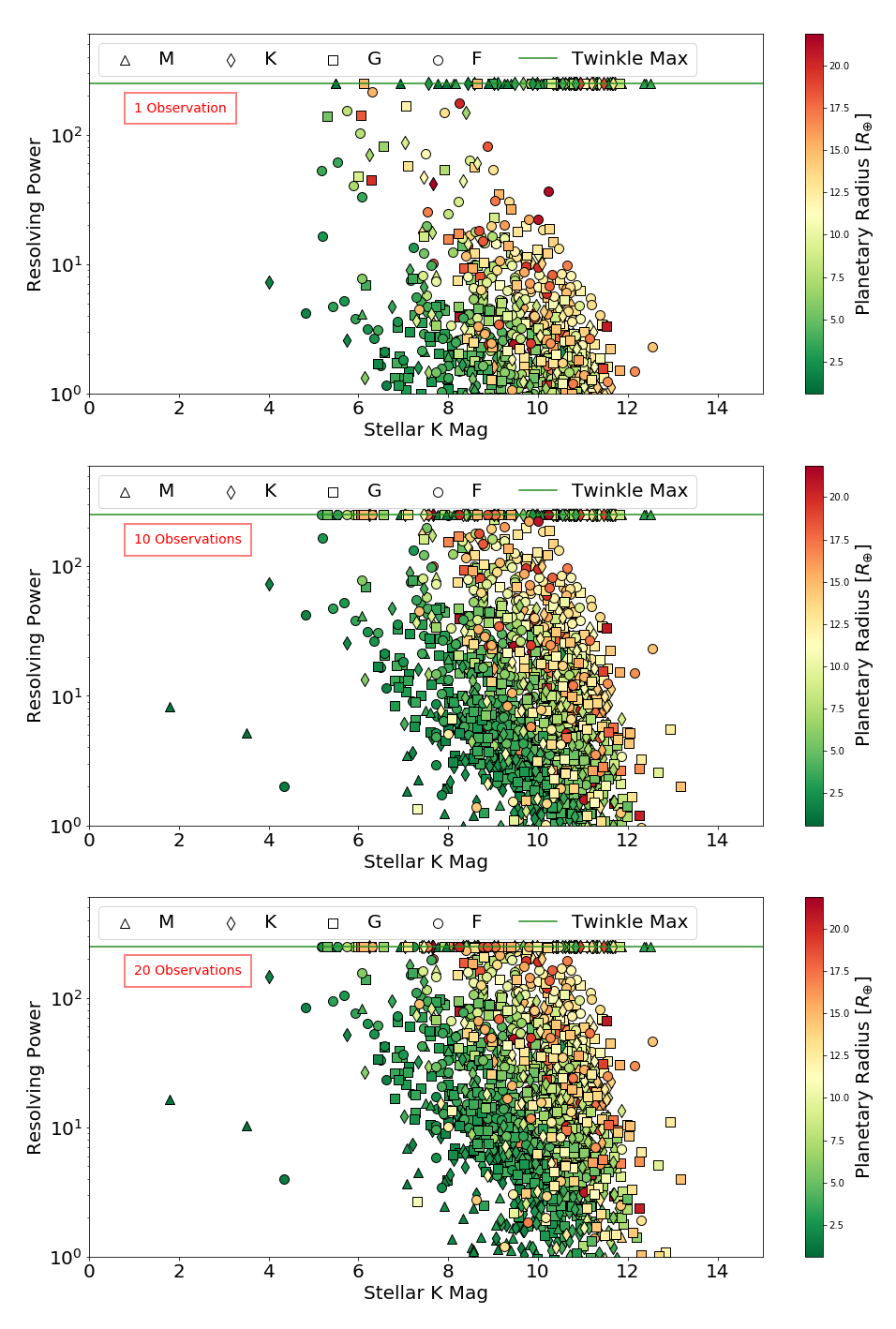}
  \caption{Achievable resolving power in the visible channel (0.4 - 1.0$\mu$m) for currently-known exoplanets and predicted TESS planets within Twinkle's field of regard assuming a requirement of SNR $\geq$ 7 for a given number of transit or eclipse observations. The colour of the point indicates the planetary radius, the shape indicates the stellar type of the host star and the green line represents Twinkle's maximum resolving power in the channel} \label{kmag ch0}
\end{figure}

\begin{figure}[ht!]
  \centering
  \includegraphics[width=0.85\textwidth]{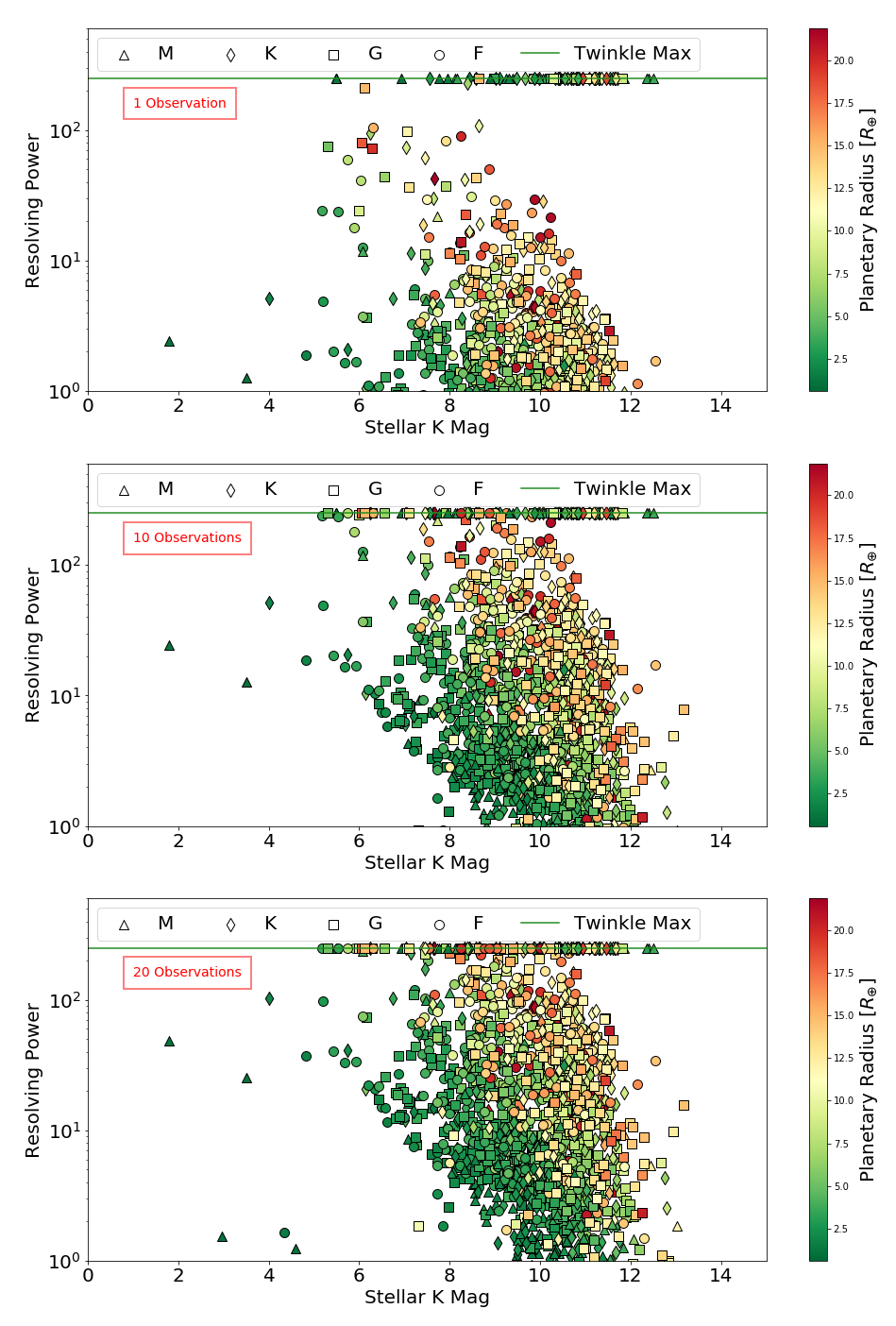}
  \caption{Achievable resolving power in the first infrared channel (1.3 - 2.42$\mu$m) for currently-known exoplanets and predicted TESS planets within Twinkle's field of regard assuming a requirement of SNR $\geq$ 7 for a given number of transit or eclipse observations. The colour of the point indicates the planetary radius, the shape indicates the stellar type of the host star and the green line represents Twinkle's maximum resolving power in the channel} \label{kmag ch1}
\end{figure}

Simulations shown here indicate that, for very bright targets, Twinkle will be capable of spectroscopy with resolutions R \textgreater 20 and sufficient precision to probe the major species present in their atmospheres and constrain lower abundance molecules. For some planets, such as HD 209458 b, this will be achievable in one transit or eclipse whilst others may require between 10 to 20 observations. The achievable resolution is plotted as a function of stellar k magnitude (Figures \ref{kmag ch0} - \ref{kmag ch1}, Appendix A) and show a cut-off limit of K mag $\sim$12 for spectroscopy (R\textgreater20). Hence many Kepler planets orbit stars too faint for spectroscopic follow-up with Twinkle but TESS is expected to provide a multitude of suitable targets. The majority of planets that can be observed spectroscopically with Twinkle are hot, giant planets or Neptunes although some smaller or cooler planets around very bright stars might be feasible. 

By simultaneously providing spectroscopic data over a large wavelength range, Twinkle will be able to reduce the degeneracies that affect current observations with Hubble and Spitzer. Whilst the spectral range of WFC3 is ideal for detecting water vapour and characterising clouds, Twinkle's wavelength range contains absorption features from a wide variety of molecules (e.g. H$_2$O, CH$_4$, CO$_2$, CO, NH$_3$, HCN, TiO, ViO). The instrument spectral resolving power allows in principle the detection of many trace gases, the only limiting factor being the precision achieved which depends upon the integration time (i.e. the number of transit or eclipse events). Twinkle’s observations will be demand-based and thus the requirements in terms of spectral resolution and SNR will depend on the user’s preferences. The values here utilised for retrievals have been selected to provide an overview of the capabilities of the satellite and to allow a potential user to assess Twinkle’s suitability for the observations they desire. From the spectral retrievals conducted here it is found that the major constituents of an atmosphere could potentially be recovered with low resolution spectroscopy (R$\sim$20). However, for weaker molecular transitions or to retrieve trace gas abundances more accurately, longer integration time may be required (e.g. Figure \ref{55cncespec}). In the case of Figure \ref{55cncespec}, the underlying issue may be the cross-section utilised for C$_2$H$_2$ which has been determined at Earth-like temperatures, not the $\sim$2000K assumed here for 55 Cnc e. This highlights the need for accurate line lists over a wide temperature range.

The visible part of the spectrum can be utilised to measure the planetary albedo, Rayleigh scattering and detect/characterise clouds. Star spots and faculae may affect the observed transit depth at wavelengths shorter than 2$\mu$m and Twinkle's spectral coverage should allow stellar activity to be monitored and to remove, or mitigate, its impact on the observations. For hotter planets, metallic resonance line (e.g. Na and K) dominate the opacities over visible wavelengths \cite{encrenaz}. The spectral resolving power considered for Twinkle in the optical (R$\sim$250) will allow for such detections on planets orbiting very bright stars. A wide spectral range will also be advantageous in the search for condensates or hazes with many species expected to condense in exoplanetary atmospheres, as suggested by current observations \cite{tsiaras_30planets,sing}.

In the thermal regime, usually probed in the infrared through eclipse observations, redundancy in molecular detection is also necessary to allow for the retrieval of the vertical thermal structure and molecular abundances (e.g. \cite{goody,tinetti_spec}). The capability to observe multiple absorption bands of the same molecule provides some redundancy and significantly improves the reliability of a detection. Additionally, insights into the vertical distribution of species can be gained by observing bands of different intensity which probe different atmospheric levels. 

For a more refined list of targets and performances, several factors not studied in this paper must also be accounted for. These are detailed in Sections \ref{obscuration} - \ref{future detections} below.

\subsection{Complimentary with Other Facilities}

Twinkle could be utilised to provide preliminary observations of targets for other observatories such as JWST and ARIEL, refining transit times and reducing the risk of missing transits due to poor ephemeris data. Given its capabilities, JWST time will be extremely precious \cite{cowan}, with over-subscription likely to be an issue, and thus any insight about the atmospheres of recently discovered planets will be highly advantageous to guide the selection of the very best targets. ARIEL will be launched in 2028 and aims at observing a very large population of exoplanet atmospheres \cite{tinetti_ariel}. A key decision for ARIEL is the selection of optimal and diverse targets before its launch and Twinkle could be utilised to inform these decisions as well as provide initial insights into the mission's science objectives. ARIEL will require a robust and efficient schedule to observe a large population of exoplanets. Hence providing constraints on the planetary, stellar and orbital parameters with Twinkle would enhance the mission's scientific yield.

Additionally, Twinkle could enhance ground-based observations. Ground-based surveys are capable of extremely high resolution spectroscopy over narrow wavebands, but the spectral continuum is unknown. By observing the same target over a broader wavelength range from space, at a lower resolution, Twinkle will be able to provide the missing, highly complementary information.

\subsection{Earth Obscuration} \label{obscuration}
The radiometric model used calculates the SNR achieved based upon observing a full transit or occultation. However, due to Twinkle’s low Earth orbit, it will not always be possible to view the transits (or eclipses) in their entirety; instead, partial transits (or eclipses) will often be observed. Observing partial transits/eclipses will reduce the SNR achieved for a given number of observations and thus the number of transits (or eclipses) predicted by the radiometric model is, in these cases, an underestimate of the number that will need to be observed. A subsequent paper will assess the impact of Earth obscuration and the effect on the number of potentially observable targets.

\subsection{Scheduling}
The observability of a target has been determined by assuming that a given number of complete transits or eclipses could be viewed during the mission lifetime. In addition to Earth obscuration, scheduling constraints, such as telescope housekeeping, slewing between targets and observations of other targets, will impact the number of transits or eclipses that are observable in a given time period. Such constraints have not been included in this exercise and the development of an optimised schedule is dependent upon an improved understanding of the additional observations required due to Earth obscuration.

The expectation is that the main issue may be observation overlaps (i.e. two planets transiting at the same time) rather than an insufficient number of potential observations. Our analysis finds that, for currently-known planets within Twinkle’s field of regard, around 85\% of targets have at least 5 transits/eclipses which could be viewed in a year whilst over 60\% of targets have 10 potential observations or more per year. Therefore, during the mission lifetime, most planets will have many transits and eclipses which could be observed. Further target selection studies will occur in due course and will incorporate these constraints.

\subsection{Future Planet Discoveries} \label{future detections}
The main focus of this work is on Twinkle’s capability to observe currently-known planets whilst also considering the predicted TESS yield. Additionally to TESS, by the launch date, other space-based missions such as Gaia, CHEOPS and K2, as well as ground-based surveys including NGTS, ESPRESSO, and WASP, are expected to have discovered hundreds of new planets around bright stars within Twinkle’s field of regard. This will provide an expanded list of planets including many that are anticipated to be observable by Twinkle.

Predicted TESS detections were included in this study to provide an indication of the number, and type, of planets projected to be discovered which will be suitable for observations with Twinkle. We find that discoveries from the TESS survey could more than double the number of exoplanet atmospheres that Twinkle is capable of observing. As further planets are discovered they will be added to this analysis to produce a comprehensive Twinkle target list.

\section{Conclusion}
In this study we present an initial survey of Twinkle’s capability for optical and infrared observations of exoplanets. For the several hundred currently-known planets which lie within Twinkle’s field of regard and whose parameters are known, we have estimated the spectral resolution which could be obtained for a given number of transit or eclipse events.

Within a single transit or eclipse observation, it is predicted that 82 existing targets could be observed in at R \textless 20 in channel 1 (1.3 - 2.42$\mu$m) whilst 68 planets could be observed spectroscopically (R \textgreater 20) with 10 transits or eclipses. The planets observable spectroscopically are found to be generally hot, with planetary radii greater than 10 R$_{\oplus}$, and orbiting bright stars (K magnitude\textless11).

Spectral retrieval simulations of HD 209458 b, GJ 3470 b and 55 Cnc e highlight the expected capability of Twinkle for atmospheric characterisation in case of planets around very bright stars. We find that most abundant molecular species, cloud and atmospheric key parameters can be retrieved reasonably well at the spectral resolution obtainable with Twinkle.

Future surveys will reveal thousands of new exoplanets, some of which will be located within Twinkle’s field of regard. Analysis of the predicted detections suggests the number of exoplanets, and exoplanet atmospheres, Twinkle is capable of characterising will dramatically increase from planets found with TESS.

\newpage

\textbf{Appendix A: K Magnitude Limits}

\begin{figure}[ht!]
  \centering
  \includegraphics[width=0.85\textwidth]{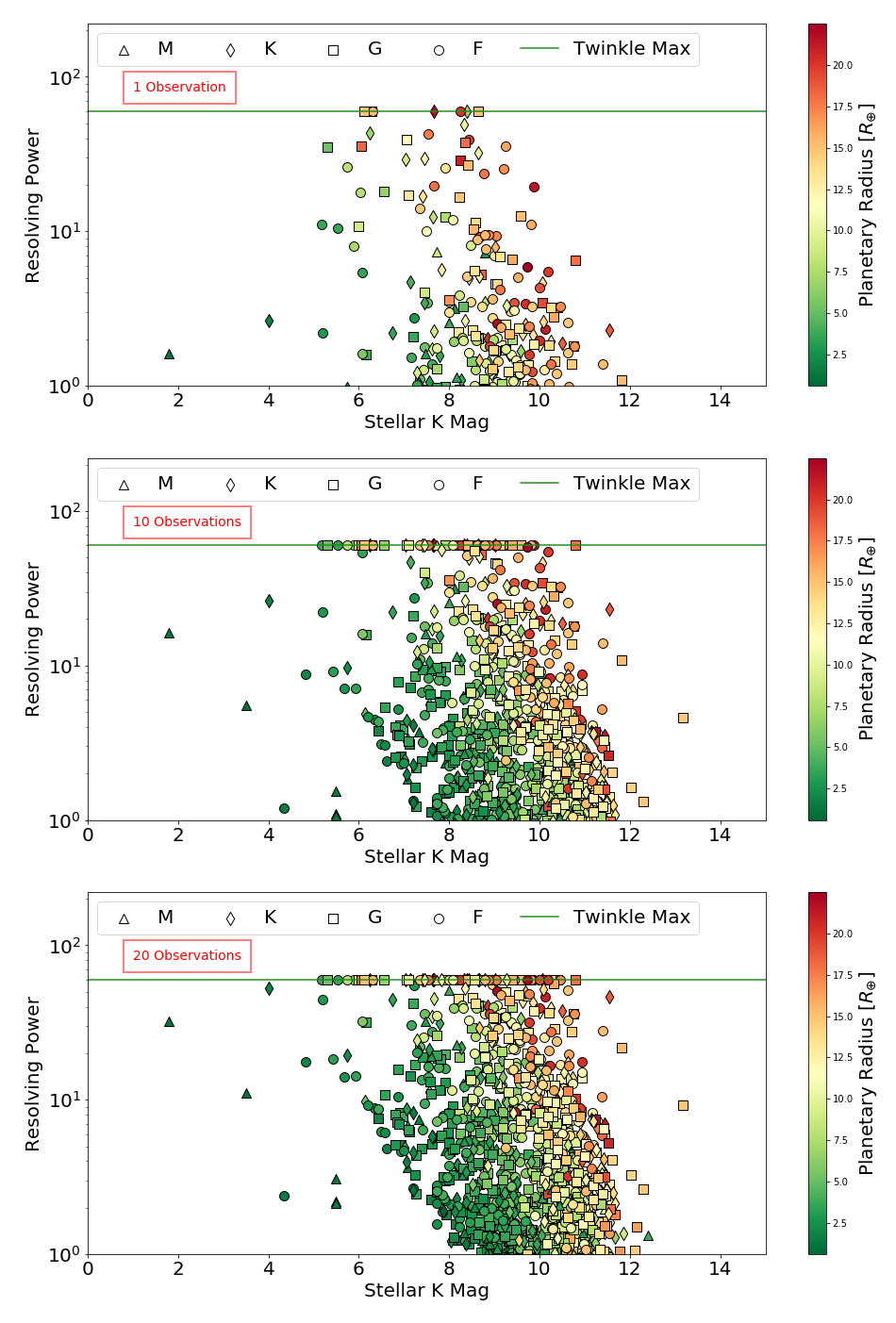}
  \caption{Achievable resolving power in the second infrared channel (2.42 - 4.5$\mu$m) for currently-known exoplanets and predicted TESS planets within Twinkle's field of regard assuming a requirement of SNR $\geq$ 7 for a given number of transit or eclipse observations. The colour of the point indicates the planetary radius, the shape indicates the stellar type of the host star and the green line represents Twinkle's maximum resolving power in the channel} \label{kmag ch2}
\end{figure}

\clearpage
\textbf{Appendix B: Retrieval Posteriors} 

\begin{figure}[ht!]
  \centering
  \includegraphics[width=1\textwidth]{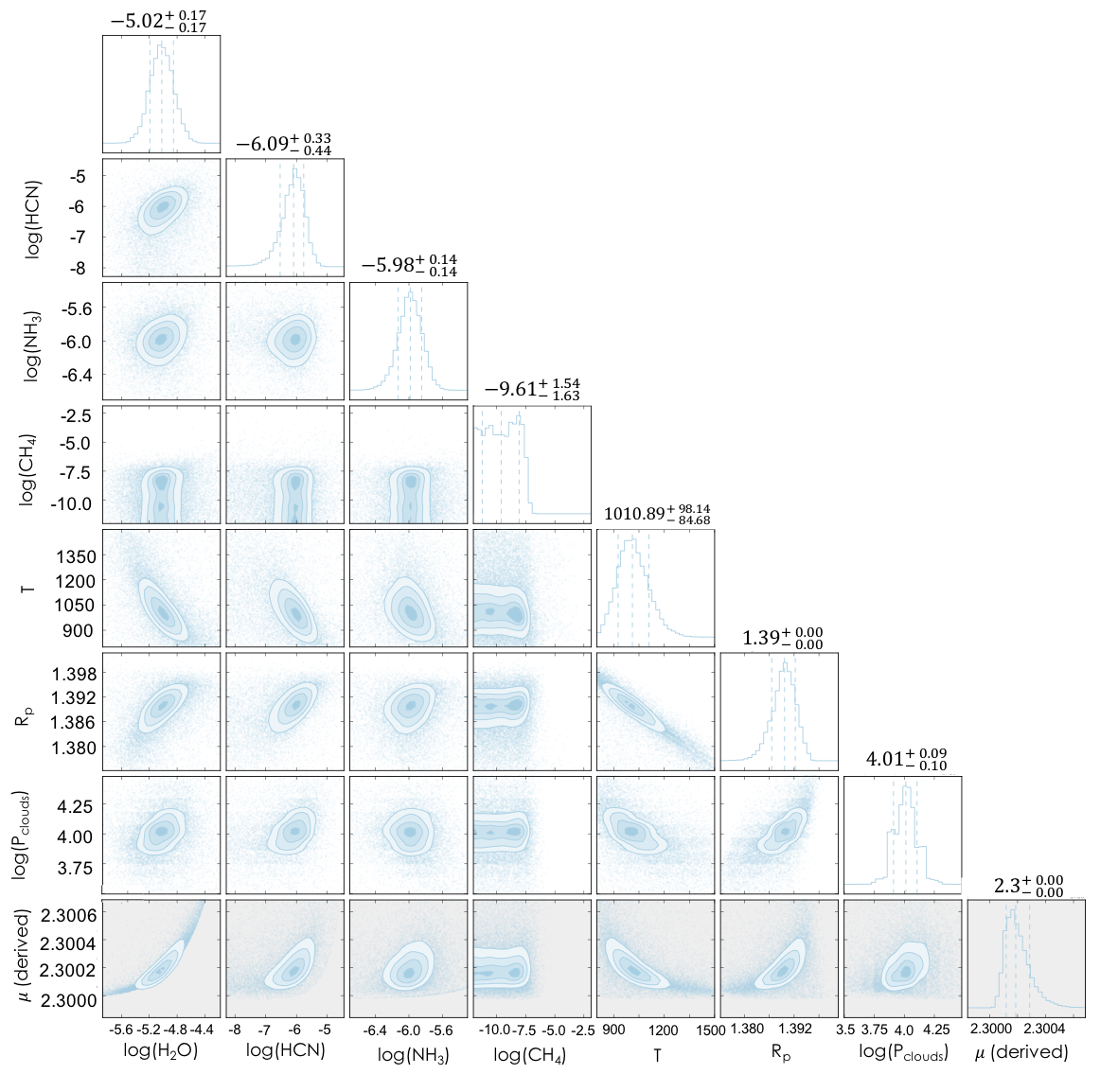}
  \caption{Posteriors for spectral retrieval of HD 209458 b (P$_{Clouds}$ = 1x10$^{4}$, H$_2$0 = 1x10$^{-5}$, HCN = 1x10$^{-6}$, NH$_3$ = 1x10$^{-6}$, CH$_4$ = 1x10$^{-8}$) at R = 250 ($\lambda$ < 2.42$\mu$m) and R = 60 ($\lambda$ > 2.42$\mu$m) with 1 transit observation which correctly recovers the major molecular abundances and cloud pressure but does not constrain CH$_4$} \label{hd209post}
\end{figure}

\begin{figure}[ht!]
  \centering
  \includegraphics[width=1\textwidth]{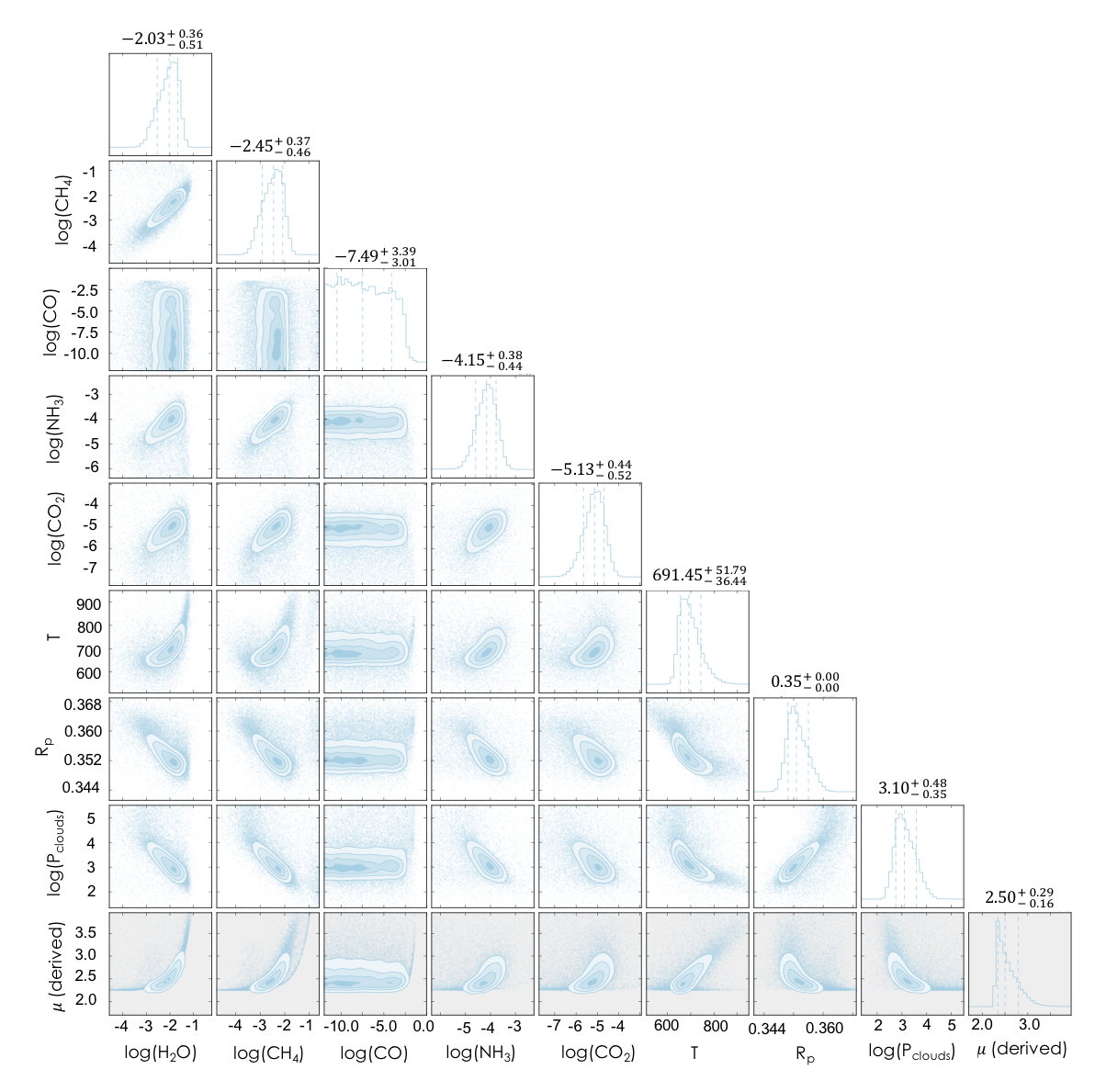}
  \caption{Posteriors for spectral retrieval for GJ 3470 b (P$_{Clouds}$ = 1x10$^{3}$, H$_2$0 = 1x10$^{-2}$, CH$_4$ = 4x10$^{-3}$, CO = 1x10$^{-3}$, NH$_3$ = 1x10$^{-4}$, CO$_2$ = 1x10$^{-5}$) at R = 65 ($\lambda$ < 2.42$\mu$m) and R = 20 ($\lambda$ > 2.42$\mu$m) with 10 transit observations which correctly recovers the major molecular abundances and cloud pressure but does not constrain CO} \label{gj3470post}
\end{figure}

\begin{figure}[ht!]
  \centering
  \includegraphics[width=1\textwidth]{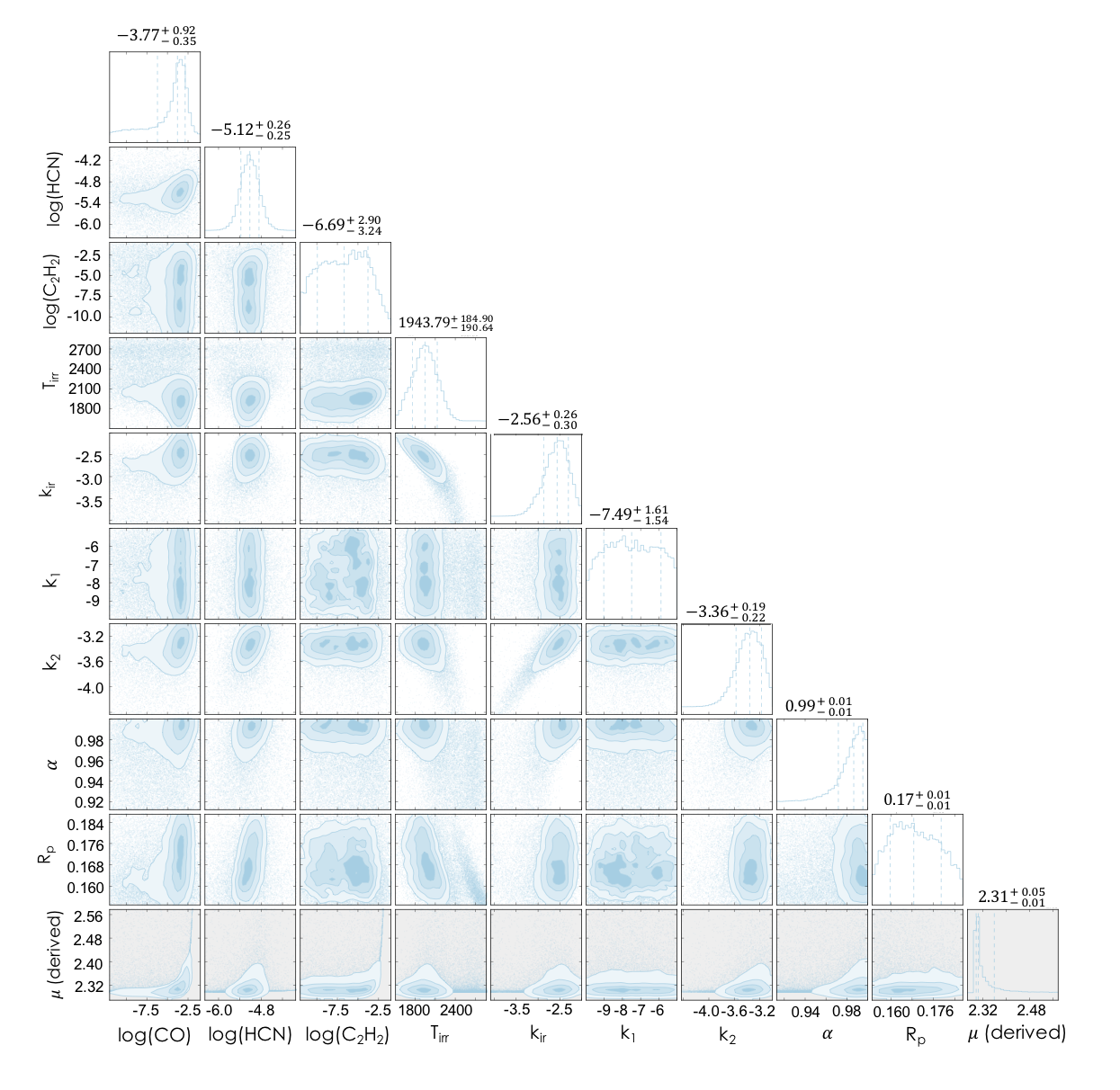}
  \caption{Posteriors for spectral retrieval for 55 Cnc e (cloud free, CO = 1x10$^{-3}$, C$_2$H$_2$ = 1x10$^{-5}$, HCN = 1x10$^{-5}$) at R = 10 ($\lambda$ < 2.42$\mu$m) and R = 20 ($\lambda$ > 2.42$\mu$m) with 10 eclipse observations which correctly recovers the HCN and CO abundances but does not obtain the correct C$_2$H$_2$ abundance} \label{55cncepost}
\end{figure}

\clearpage

\bibliographystyle{plain}
\bibliography{main}

\end{document}